\documentclass[11pt,a4paper]{article}
\pdfoutput=1
\usepackage[utf8x]{inputenc} 
\usepackage{jheppub}
\usepackage{amsfonts}
\usepackage{amsmath}
\usepackage{slashed}
\usepackage{amssymb}
\usepackage{graphicx}
\usepackage{epic}
\usepackage{eepic}
\usepackage{epsfig}
\usepackage{latexsym}
\usepackage{color}
\usepackage{float}
\usepackage{multirow}
\usepackage{hyperref}
\usepackage{enumitem}
\hypersetup{colorlinks=true,citecolor=red,linkcolor=magenta,urlcolor=blue}
\usepackage[caption=false]{subfig}
\usepackage{natbib}
\usepackage{relsize}
\usepackage{breqn}
\usepackage{shorthand}
\usepackage{subfig}
\usepackage[normalem]{ulem}

\def\dis{\displaystyle}
\def\beq{\begin{equation}}
\def\eeq{\end{equation}}
\def\barr{\begin{array}}
\def\earr{\end{array}}

\definecolor{darkred}{cmyk}{0,1,1,0.4}

\usepackage{color}
 \usepackage[normalem]{ulem}
\long\def\/*#1*/{}
\usepackage{color}
 \usepackage[normalem]{ulem}
\definecolor{darkgreen}{cmyk}{1,0,1,0.4}
\definecolor{darkred}{cmyk}{0,1,1,0.4}

 \def\hldc#1{\textcolor{darkgreen}{\large\textsl{#1}}}

\title{Neutrino and $Z'$ phenomenology in an anomaly-free $\mathbf{U}(1)$ extension: role of higher-dimensional operators}

\author[a]{Debajyoti Choudhury,}

\author[a]{Kuldeep Deka,}
\emailAdd{kuldeepdeka.physics@gmail.com}

\author[b,c]{Tanumoy Mandal,}

\author[a]{Soumya Sadhukhan}

\affiliation[a]{Department of Physics and Astrophysics, University of Delhi, Delhi 110 007, India}
\affiliation[b]{Indian Institute of Science Education and Research Thiruvananthapuram, Vithura, Kerala, 695551, India}
\affiliation[c]{Department of Physics and Astronomy, Uppsala University, Box 516, SE-751 20 Uppsala, Sweden}

\date{\today}

\abstract{We consider an anomaly-free $\mathrm{U}(1)$ extension of the Standard Model with
  three right-handed neutrinos (RHNs) and two complex scalars, wherein
  the charge assignments preclude all tree-level mass terms for the
  neutrinos.  Considering this setup, in turn, to be only a low-energy
  effective theory, we introduce higher-dimensional terms {\em a la}
  Froggatt-Nielsen to naturally generate tiny neutrino masses. One of
  the RHNs turns out to be very light, thereby constituting the main
  decay mode for the $Z'$ and hence relaxing the LHC dilepton
  resonance search constraints. This very RHN has a lifetime comparable to or bigger
  than the age of the Universe, and, hence, could account for a
  non-negligible fraction of the dark matter.}

\begin{document}

\maketitle
\def\lapp{\mathrel{\rlap{\raise.5ex\hbox{$<$}}
                    {\lower.5ex\hbox{$\sim$}}}}
\def\gapp{\mathrel{\rlap{\raise.5ex\hbox{$>$}}
                    {\lower.5ex\hbox{$\sim$}}}}    

\section{Introduction}

Of all the lacunae besetting the Standard Model (SM), the existence of
Dark Matter (DM) and the flavour problem are rather vexing ones. While the
masses of the charged fermions as well as the mixing amongst the
quarks can be explained by postulating a certain set of Yukawa
couplings, the large hierarchy between these is a rather disquieting
feature. Over the last few decades, several disparate sets of
theories have been proposed including (but not limited to) ($i$)
additional symmetries, discrete or continuous, gauged or global~\cite{Mohapatra:1980yp,Lazarides:1980nt,Fritzsch:1999ee,Babu:2002dz,Strumia:2006db,Ma:2006sk,Altarelli:2010gt,King:2013eh,Ma:2015mjd,PhysRevLett.43.92,PhysRevD.29.1504},
($ii$) quark compositeness, whether in terms of further constituents,
bound together by some unspecified force~\cite{Strassler:1995ia,Hayakawa:1997ud,Haba:1997bj,Haba:1998wf} or, in the more modern parlance, in
terms of a higher-dimensional theory, often with a nontrivial
gravitational background~\cite{Dienes:1998sb,ArkaniHamed:1998vp,ArkaniHamed:1999dc,Grossman:1999ra,Huber:2000ie}. Very often, though, such efforts are faced
with undesirable phenomenological consequences (unsuppressed
flavour-changing neutral currents being one such), and these issues
can be cured only through the introduction of further
complications. Even more damagingly, no corroborating evidence has
been found in terms of additional particles (that many such scenarios
posit) or interactions.

The situation has been exacerbated in recent years by the observation
of neutrino oscillations~\cite{Fukuda:1998mi,Ahmad:2001an,Abe:2011fz,An:2012eh,Ahn:2012nd} and these seemingly call
out for nonzero neutrino masses. Indeed, the consequent mixing angles
in the neutrino sector are quite well-determined and so is
one difference in the squares of masses~\cite{Tanabashi:2018oca,Esteban:2018azc}. For the other independent
difference, only the magnitude (and not the sign) is known and also
unknown are the nontrivial phases that are possible in the mixing
matrix. And while the oscillation data is only sensitive to the
difference in mass-squareds and not the absolute mass scale, the
latter is very-well constrained to $\sum m(\nu_i) <
(0.340-0.715)$ eV~\cite{Tanabashi:2018oca} -- where $\nu_i$ are (cosmologically) stable light
neutrinos -- from a host of cosmological data.
%% (including the cosmic microwave background radiation, 
Interestingly, direct bounds from terrestrial experiments (such as 
decays) are significantly weaker \cite{Aker:2019uuj}. 

Of course, neutrino masses (absent within what is known as the SM) can
be trivially obtained by introducing right-handed neutrino (RHN) fields
$\nu_{jR}$ and ascribing additional Yukawa terms. Tuning the said
couplings so as to obtain the requisite masses and mixings is a
seemingly trivial task, once a further hierarchy in the Yukawa
couplings (necessitated by the smallness of the neutrino masses) is
accepted.  The pitch is further queered, though, by the fact that with
the RHNs being gauge singlets, terms such
$\overline{(\nu_{jR})^c} \nu_{kR}$ are gauge invariant, and
being unprotected by any symmetry\footnote{Lepton number
  conservation is only accidental within the SM and is, actually,
  broken by nonperturbative effects.}, can be arbitrarily large. Beset
with such large Majorana masses, the $\nu_{jR}$ can be integrated out
from the low-energy theory, leaving the SM neutrinos with tiny
masses. Indeed, this very observation led to a cure
in the form of the {\em seesaw mechanism}, wherein a large (dynamical)
scale is set for the Majorana mass matrix $m_M$, with
and the usual Dirac mass matrix ($m_D$) for the neutrinos being
unsuppressed, so that on integrating out the heavy fields, the light
fields are left with an effective mass matrix $m_{\rm eff} \sim m_D^T
m_M^{-1} m_D$ which, on diagonalization, should yield the
observed masses and mixings.

With $m_D$ being proportional to the vacuum expectation value of the
SM Higgs $H$, the aforementioned structure could have been divined in
an effective theory.  Written in a gauge-invariant form, the Weinberg
operator~\cite{Weinberg:1979sa} reads $c_{ij} \overline{L_i^c} L_j H
H/\Lambda$ where $L_i$ are the left-handed lepton doublets and
$\Lambda$ is the cutoff scale (perhaps allied to the largest
eigenvalue of $m_M$ above). The dimensionless constants $c_{ij}$
constitute a symmetric matrix and can be thought of as parametrizing
the structure of $m_D^T m_M^{-1} m_D$. Once again, choosing
$c_{ij}/\Lambda$ to be small enough and ascribing the necessary
structure, the correct set of masses and mixings can be obtained.

All of the aforementioned mechanisms to generate the light neutrino
masses ``suffer'' from the requirement of either postulating very tiny
couplings or a very large scale ({\em e.g.}, $m_M \sim 10^{11}$ GeV
for the so-called ``type-I'' seesaw mechanism) rendering such theories
essentially untestable by current experiments. It would, thus, be very
attractive to have a theory for neutrino masses with a characteristic
scale ${\cal O}$(10 TeV) or lower so that it is testable at the LHC,
the $B$-factories {\em etc.}. Examples are
  scenarios~\cite{Deppisch:2015qwa,Huitu:2008gf} of TeV scale
  RHNs with a significant mixing with the SM $\nu$s achieved through
  the realization of low scale seesaw through non-trivial flavor
  structure. Similarly, models with radiative neutrino mass
  generation and/or inverse seesaw~\cite{PhysRevD.88.113001} also exist; 
  with relatively light RHNs, these can
  be probed at colliders. Looking for well-motivated scenarios 
  that incorporate experimentally testable RHNs in a more
  broader scheme, is the goal we set for this work.  

Before we start this in right earnest, it is worthwhile to remind
ourself of a particularly elegant proposal addressing the fermion mass
hierarchy. As Froggatt and Nielsen (FN)~\cite{Froggatt:1978nt} pointed out,
ascribing the Higgs and quark fields with some extra charges
(corresponding to a discrete or a continuous symmetry) would, in
general, render the usual Yukawa terms untenable. Instead, higher
dimension terms could be written by inserting an appropriate number of
a ``flavon'' scalar field ${\cal F}$. The choice of the charges would
dictate the powers of the ratio $\langle {\cal F} \rangle / \Lambda$
($\Lambda$ being the cutoff scale) in the effective mass terms and
hence their scales. Apparently, then, this simple ruse can lead to
correct masses and mixings without the need for imposing a large
hierarchy in the Yukawa couplings~\cite{Babu:2009fd}; and the non-renormalizable nature of
the theory could be explained as being the result of integrating out
unspecified fields, the nature of which would depend on the particular
ultraviolet completion of the FN-scenario.
Unfortunately, though, the simplest such models turn out to
be phenomenologically inconsistent, failing to satisfy the constraints
from rare decays while still explaining the masses and the mixings.

While we would not discuss charged fermion masses in this paper, it is
still instructive to examine the FN mechanism and, in particular,
where the flavor charges correspond to a $\mathrm{U}(1)$ symmetry. An extra
gauged $\mathrm{U}(1)$ can, of course, appear in many a scenario, ranging from
flavor models to theories of compositeness to GUTs
\cite{Marshak:1979fm,
Mohapatra:1980qe,
Baek:2001kca,
Khalil:2006yi,
Iso:2009ss,
Khalil:2010iu,
Chao:2010mp,
Heeck:2011wj,
Das:2013jca,
Altmannshofer:2014cfa,
Baek:2015mna,
Biswas:2016yan,
Biswas:2017tce,
Singirala:2017see,
Asai:2017ryy,
Arcadi:2018tly,
Kamada:2018zxi,
Banerjee:2018eaf,
Jana:2019mez,
Nam:2019wjs,
Nam:2019zaa}.
Naturally, the
phenomenological consequences are very well studied~\cite{Erler:1999ub,Langacker:2008yv,Basso:2008iv,Erler:2009jh,Salvioni:2009mt,Salvioni:2009jp,Ekstedt:2016wyi,Bandyopadhyay:2018cwu,Aebischer:2019blw,Dudas:2013sia,Okada:2018tgy,Deppisch:2019ldi} and strong
constraints (in the mass--gauge coupling plane) emerge from a variety
of measurements, ranging from rare decays, anomalous magnetic moments
of the electron or muon, electroweak precision tests (performed at
the $Z$-peak) to direct observation at the LHC. The relative strengths
of the constraints are determined by the $\mathrm{U}(1)$ charge assignments for
the light SM fermions. The latter, of course, are not entirely
arbitrary as the $\mathrm{U}(1)$ extension needs to be anomaly-free. In
particular, some of the strongest constraints emanate from the lack of
unexplained, yet discernible, peaks in the invariant mass spectra for
dijet or dilepton production at the LHC. 

In this paper, we pursue a modest goal. Starting with an anomaly-free
$\mathrm{U}(1)$ extension of the SM (augmented by the mandatory three
RHN fields), we employ a Froggatt-Nielsen-like
mechanism, but restricted strictly to the neutrino sector. We find
that $(a)$ the neutrino masses and mixings can be explained with very
moderate choices for the Yukawa couplings and a $\mathrm{U}(1)$ scale of a few
TeVs; ($b$) simultaneously, the dilepton branching fraction of the
$Z'$ is suppressed so that even with very natural choices of
parameters, a $Z'$ as light as 3 TeV is perfectly consistent with the
LHC results and, yet, ($c$) novel signatures are predicted at the LHC.

An additional bonus is the natural emergence of a viable Dark Matter
candidate, thereby addressing the second (and, perhaps, even more
pressing) lacuna of the SM that we had alluded to. In fact, as many as
two ultralight particles (a pseudoscalar and a RHN)
appear in the spectrum, with their masses uplifted only by
higher-dimensional operators. While not strictly stable, the lighter
of these has a lifetime comparable to or greater than the age of the
universe.  Being charged under the extra $\mathrm{U}(1)$ (although neutral
under the SM gauge group), their interactions are large enough to be
interesting in the context of both cosmology and direct detection.

The rest of the paper is organized as follows. We discuss the anomaly-free $\mathrm{U}(1)_z$ extension model in Section~\ref{sec:u1_extn} including gauge, scalar and fermionic sectors
of the model. We devote the phenomenology of the neutrino sector and $Z'$ boson in Sections~\ref{sec:numass} and \ref{sec:zppheno} respectively wherein we discuss various exclusion limits. In Section~\ref{sec:DM}, we explore the possibility whether the new ultralight particles present in our model can act as a suitable dark matter candidate. Finally, we summarize and conclude in Section~\ref{sec:sumcon}.

\section{The $\mathbf{U(1)_z}$ extension}
\label{sec:u1_extn}

The gauge sector of the SM
(i.e. $\mathrm{SU}(3)_c\times\mathrm{SU}(2)_L\times\mathrm{U}(1)_Y$)
is extended by a new $\mathrm{U}(1)_z$ gauge symmetry with the
associated gauge coupling $g_z$. Presence of multiple $\mathrm{U}(1)$s
in a gauge theory can, in general, lead to kinetic mixing. However, it
is always possible to rotate away this kinetic mixing at a given scale
(due to running of the couplings, it can be regenerated at other
scales). In this paper, we are only interested in the effective
TeV-scale phenomenology and therefore, we 
make the simplifying assumption
that the kinetic mixing between $\mathrm{U}(1)_Y$ and
$\mathrm{U}(1)_z$ is removed by a suitable field rotation at the
TeV-scale. 
%\sout{Note that, one can, in general, start with
%  $\mathrm{U}(1)_1\times\mathrm{U}(1)_2$ and then break it down to
%  $\mathrm{U}(1)_Y$ at a high scale and proceed with the electroweak
%  symmetry breaking (EWSB) as usual. But, it turns out that, it is
%  always possible to make $\mathrm{U}(1)_1\times\mathrm{U}(1)_2$ looks
%  like $\mathrm{U}(1)_Y\times\mathrm{U}(1)_z$ by suitable redefinition
%  of the gauge fields and rescaling the gauge couplings.} \comment{Is
%  the preceding crossed-out part necessary?}  
We further assume that
the SM fields are charged under the new $\mathrm{U}(1)_z$ and the
corresponding quantum number of a field $\mc{F}$ being denoted by
$z_{\mc{F}}$.  With the SM fields too allowed to have nonzero
$z_{\mc{F}}$, anomaly cancellation is a concern. Postponing this
concern until later, we begin by considering the bosonic sector.

\subsection{Scalars and symmetry breaking}
At a scale much higher than the electroweak symmetry breaking (EWSB) scale,
the $\mathrm{U}(1)_z$ is broken by one or more SM singlets $\chi_A$ carrying 
charges $z_{\chi_A}$. While a single $\chi$ suffices for 
the requisite symmetry breaking, we generalize the situation for 
phenomenological reasons which will be clear later. 
The Lagrangian for the scalar sector is given by
\begin{equation}
\mathcal{L} \supset \lt(D^\mu H\rt)^\dagger\lt(D_\mu H\rt) + 
            \sum_A \lt(\widetilde{D}^\mu \chi_A\rt)^\dagger \lt(\widetilde{D}_\mu \chi_A\rt)
             - V(\{H^\dagger H\}, \{\chi_A^\dagger \chi_A\})\ ,
\end{equation}
where $H$ denotes the SM Higgs doublet and our assumption that the charges
$z_{\chi_A}$ are such that trilinear terms in the potential are not admissible would be
vindicated later. The covariant derivatives, for a generic field $\mc{F}$ is, of course, given by
\begin{equation}
D_\mu = \partial_{\mu} - i g_s T^a G^a_\mu - 
ig_w \frac{\sigma^i}{2} W^i_\mu - ig_y \dfrac{Y}{2} B_\mu -ig_z 
  \dfrac{z_{\cal F}}{2} X_{\mu}\ ;~~~\widetilde{D}^\mu = \partial_{\mu}-ig_z 
  \dfrac{z_{\cal F}}{2} X_{\mu}\ ,
\end{equation}
with $X_\mu$ being the new gauge boson. The second, third and the
fourth terms on the r.h.s. of the first of
  the above equations correspond to the $\mathrm{SU}(3)_c$,
$\mathrm{SU}(2)_L$ and $\mathrm{U}(1)_Y$ gauge groups of the SM with
gauge couplings $g_s$, $g_w$ and $g_y$ respectively.

The symmetry is broken, in two steps, by the vacuum expectation values
(vevs) of $\chi_A$ and $H$ fields.
\begin{equation}
\langle \chi_A \rangle \equiv \frac{x_A}{\sqrt{2}} \ , \qquad
\langle H \rangle \equiv \left(0 \quad \frac{v_h}{\sqrt{2}} \right)^T \ .
\end{equation}
We assume that, in case multiple $\chi_A$ are invoked, the
corresponding vevs $x_A$ are of the same order, i.e., no hierarchy is
introduced between them. Furthermore, for the sake of simplicity, we
do not admit spontaneous $CP$ violation, or in other words, nontrivial
phases between the vevs.

While the expression for $W$-boson mass $M_W$ remains unchanged, the (mass)$^2$ 
matrix for the neutral gauge bosons is now modified to
\begin{equation}
\mathcal{M}^2= \dfrac{1}{4} 
  \left[
  \barr{ccccc}
           g_y^2v_h^2 &\qquad& -g_yg_wv_h^2 & \qquad & g_yg_z z_Hv_h^2 \\[1ex]
           -g_yg_z v_h^2 & & g_w^2v_h^2 & & -g_wg_z z_Hv_h^2 \\[1ex]
           g_yg_z z_hv_h^2 && -g_wg_z z_Hv_h^2 && 
                     g_z^2\lt(z_H^2v_h^2+\displaystyle\sum_A z_{\chi_A}^2 x_A^2\rt)
  \earr                            
  \right].
\end{equation}
Clearly, $\mathcal{M}^2$ is a rank-2 matrix, and can be diagonalized by an orthogonal 
matrix $\mc{O}$ defined through
\begin{equation}
\begin{bmatrix}
A_\mu \\
Z_\mu \\
Z^\prime_\mu  
\end{bmatrix} \ = \ \begin{bmatrix}
\cos w & \sin w & 0 \\
-\cos t \ \sin w & \cos t \ \cos w & \sin t \\
\sin t \ \sin w & -\sin t \ \cos w & \cos t
\end{bmatrix}\ \begin{bmatrix}
B_\mu \\
W_{3\mu} \\
X_\mu 
\end{bmatrix}
\equiv \mc{O}^\dagger \, \begin{bmatrix}
B_\mu \\
W_{3\mu} \\
B^\prime_\mu 
\end{bmatrix}.
\end{equation}
\\
The Weinberg angle $w$ remains unaltered, namely
$w = \tan^{-1} (g_y / g_w)$
whereas for $e= g_w \sin w$, the $Z\leftrightarrow Z^\prime$ mixing angle $t$ is given by
\begin{equation}
\dfrac{4e z_Hg_z}{\sin 2w} \, \cot 2t 
     = \frac{g_z^2}{v_h^2}\Big(\sum_A z_{\chi_A}^2 x_A^2+z_H^2 v_h^2\Big) 
      - \dfrac{4 e^2}{\sin^22w}\ .
\end{equation}
%%%%%%%
The heavy neutral gauge boson masses are given by 
%%%%
\begin{equation}
\label{eq:mass}
M_{Z,Z'}^2 = \dfrac{e^2 v_h^2 \cos^2 t}{\sin^2 2w} 
       +\frac{g_z^2}{4}\left(z_H^2 v_h^2 + \sum_A z_{\chi_A}^2 x_A^2\right) 
              \sin^2 t 
        \mp \dfrac{e g_z z_H v_h^2}{2\sin 2w} \sin 2t\ .
\end{equation}
%%%%
 The shift in $M_Z$ imposes a constraint on the parameter space of the model,
which, as far as this sector is concerned, could be considered of being 
three-dimensional, namely defined by $g_z$, $z_H$ and the combination
$\sum_A z_{\chi_A}^2 x_A^2$. Note, however, that the above tree-level expression cannot 
be immediately compared with the experimentally measured $M_Z$ as quantum
corrections need to be included. We return to this point later.

Another example of such changes would be that wrought by the
scalar sector. The very inclusion of a single $\chi_A$ field results,
after the breaking of the $\mathrm{U}(1)_z$ and in the unitary gauge, in an
additional scalar field. Although a SM singlet, this can mix with the
SM Higgs field (owing to terms such as $H^\dagger H \chi^\dagger
\chi$) resulting in two physical scalars $h_{1,2}$. It thus needs to
be ensured that at least one of the two eigenstates has a mass of 125
GeV and couplings (both gauge and Yukawa) not significantly different
from the SM Higgs. As it would turn out, this holds almost trivially for 
the parameter space that we are interested in. It should also be apparent that 
once one of these (say $h_1$) is forced to be very SM-like, the other ($h_2$), 
being singlet-dominated, would have very small production cross sections at the 
LHC. The modes of interest would be $q \bar q \to Z' h_2$ (analogous to the 
Bjorken process) and $gg \to h_1 h_2$ with ($h_2$-sstrahlung). 
Understandably, such an $h_2$ would have expected detection thus far. 

The introduction of each additional $\chi_A$ results in the physical
spectrum being enhanced by a pair of spin-0 particles, one scalar and
one pseudoscalar.  Particles in each sector could mix amongst
themselves with inter-sector mixing disallowed as long as additional
$CP$-violation (explicit or spontaneous) is not introduced in the
Higgs sector. It should be noted that this introduces a new class of
modes, namely the $Z'$ (on-shell or off-shell) going to a
scalar-pseudoscalar pair.

\subsection{Example with two $\chi$ fields}

As an example, consider the special case of there being two such
fields $\chi_1$ and $\chi_2$. This would prove to be of particular
interest in the context of neutrino mass generation. If the 
corresponding $\mathrm{U}(1)_z$ charges $z_{\chi_1}$ and $z_{\chi_2}$ are not 
integral multiples of each other\footnote{All we really need to ensure 
is that ratio of the charges be different from 1, 2 or 3.}, the most 
general form of the scalar potential invariant under $\mathrm{U}(1)_z$ is given 
by
\begin{equation}
V(\chi_1, \chi_2) = - \mu_1^2 \chi_1^\dagger \chi_1 - \mu_2^2 \chi_2^\dagger \chi_2 
                  + \frac{\lambda_1}{2}\,  (\chi_1^\dagger \chi_1)^2
                  + \frac{\lambda_2}{2}\,  (\chi_2^\dagger \chi_2)^2
              + \lambda_{12}\, (\chi_1^\dagger \chi_1)\, (\chi_2^\dagger \chi_2) \ .
\label{2chi_pot}
\end{equation}
Here, we have deliberately suppressed terms containing the SM Higgs
field.  This simplifying approximation, apart from being very good at
energies far above the electroweak scale, serves to highlight certain
salient features.  As is immediately apparent, $V(\chi_1, \chi_2)$ is
invariant under a global $\mathrm{U}(1) \times \mathrm{U}(1)$ with each factor
associated with one of $\chi_{1,2}$. The most general symmetry
breaking would, thus, result in two Goldstone fields ${\cal G}_{1,2}$. 

One linear combination of the two $\mathrm{U}(1)$s is gauged, viz.  $\mathrm{U}(1)_z
\sim z_{\chi_1} \left[\mathrm{U}(1)\right]_1 + z_{\chi_2} \left[\mathrm{U}(1)\right]_2$,
and the corresponding combination of ${\cal G}_{1,2}$ appears as the
longitudinal component of $B'_\mu$. The orthogonal combination appears
in the spectrum as a {\em massless} pseudoscalar $A$. While it may
seem that the presence of a Goldstone in the theory would render it
phenomenologically nonviable, this is not necessarily so,
as we argue below. 
% and we return to this point later.

A true Goldstone would translate to a long-range force. And, even if
it acquired a small mass through an explicit symmetry-breaking term,
such a particle would still make its presence felt in a variety of
interactions. Some of the strongest bounds emanate from low-energy
processes including, but not limited to, astrophysical ones.  Indeed,
Majoron models ({\em i.e.}, ones, analogously to us, wherein a
Goldstone arises in spontaneously breaking a lepton number symmetry in
the quest to achieve Majorana masses~\cite{Brune:2018sab} have been
well-studied in this context. As it turns out, though, in the present case, 
the Goldstone would have very suppressed couplings with the SM fermions (thanks 
to the quantum number assignments) and, consequently, such bounds are 
expected to be satisfied easily. Indeed, the only SM particle that 
it might have unsuppressed couplings with is the Higgs, thanks to possible 
terms such as 
\[
V(H, \chi_1, \chi_2) \supset H^\dagger H \, 
    \left( \lambda_{H\chi_1} \chi_1^\dagger \chi_1 
         + \lambda_{H\chi_2} \chi_2^\dagger \chi_2 \right) \ .
\]
Post EWSB, these immediately give rise
to a trilinear $HAA$ coupling, with a strength determined by 
the $\lambda_{H\chi_{1,2}}$. It is interesting to note that the 
data on the 125~GeV scalar still allows for a non-significant 
invisible decay of the particle~\cite{Sirunyan:2018owy}
and this can be used to implement an upper bound on these 
quartic couplings. 

It should be realised, though,
that the Goldstone would, in general, be lifted by quantum
corrections, rendering it a pseudo-Nambu-Goldstone Boson (pNGB). For
example, consider an effective theory exemplified by the inclusion of
higher-dimensional terms in the Lagrangian parametrizing unknown
effects emanating from physics at still higher energies. While
Eq.~\eqref{2chi_pot} represents the most general gauge-invariant
potential consistent with renormalizability, once nonrenormalizable
terms are allowed, more terms can be present.  In the next section, we
would argue for $z_{\chi_1} = -3/4, \, z_{\chi_2} = -4$, and for such
a case the lowest-dimensional term that breaks the global
$[\mathrm{U}(1)]^2$ down to $\mathrm{U}(1)_z$ is $\chi_1^{16}
\chi_2^{*3}$. Such a large engineering dimension of the operator
would, typically, generate only a very small mass for the pNGB {\em
  viz.} ${\cal O}(x_i^{16}/\Lambda^{15})$ where $\Lambda$ is the
cutoff scale. For $x_i / \Lambda \lapp 0.1$, a very reasonable
restriction, this would leave the pNGB in a milli-eV range,
reminiscent of axionic dark matter models.

It needs to ascertained, though, whether $\chi_{1,2}$ are allowed to
have gauge-invariant Yukawa terms involving any new fermions in the
theory (such as RHN fields) and whether these
explicitly break $[\mathrm{U}(1)]^2$ down to $\mathrm{U}(1)_z$. If
such be the case, the pNGB would be lifted courtesy loop
corrections\footnote{It should be obvious that, with the couplings of
  $\chi_{1,2}$ with $H$ or $B'_\mu$ preserving
  $\left[\mathrm{U}(1)\right]^2$, the corresponding corrections to
  $V(\chi_1, \chi_2)$ would not lift the Goldstone.}.

Were it desirable to substantially raise the Goldstone, it could be
trivially done through the introduction of a third singlet scalar
$\chi_3$, which, in principle, could actually increase the global
symmetry to $[\mathrm{U}(1)]^3$. On the other hand, this may allow for
renomalizable terms breaking down the global symmetry to just a single
$\mathrm{U}(1)$ to be identified with the gauged $\mathrm{U}(1)_z$.
For example, working with the previously assigned quantum numbers for
$\chi_{1,2}$, if one introduces a third scalar $\chi_3$ with a charge
$13/4$, then a term such as $(\mu_{123} \chi_1^* \chi_2 \chi_3 +
\textrm{H.c.})$ would break the symmetry softly. On the other hand, if
$z_{\chi_3} = 5/2$, then a hard breaking is achieved through
$(\lambda_{1123} \chi_1^2 \chi_2 \chi_3^* + \textrm{H.c.})$.

It is easy to see most such augmentation of the scalar sector does not
materially alter low energy phenomenology except, perhaps, to
ameliorate some issues with the evolution history of the early
universe. As far as LHC signals go, the most drastic changes would be
in the decays of the $Z'$ wrought by the proliferation of states
(three and two $\mathrm{SU}(2)$-singlet scalars and pseudoscalars).
Since these effects are easily computed and are not very germane to
the issues that we are primarily interested in, we will not discuss
such a three-singlet scenario any further.

%\subsection{Phenomenology of the spin-$0$ sector}

To start with let us work under the assumption that the $\chi$-sector
is essentially decoupled from the SM Higgs sector.  This is not too
drastic an approximation at energy scales much higher than the
electroweak scale (except as far as the decays of the $\chi_i$ into
the SM Higgs is concerned). The potential, then, is described by 
that in Eq.~\eqref{2chi_pot}. Denoting the $\chi$ fields, post symmetry breaking, by
\begin{equation}
\chi_{1,2} = \frac{1}{\sqrt{2}} \left(x_{1,2} + \xi_{1,2} + i \rho_{1,2} \right) \ ,
\end{equation}
where $\xi_{1,2}, \rho_{1,2}$ are real fields, the massless 
pseudoscalar is given by
\begin{equation}
    A = \rho_1 \, \sin\gamma_A - \rho_2 \, \cos\gamma_A \ , 
    \qquad \tan\gamma_A = \frac{z_{\chi_2} x_2}{z_{\chi_1} x_1},
\end{equation}
with the orthogonal combination being absorbed to reappear as the
longitudinal mode\footnote{With the SM Higgs not yet acquiring a
nonzero vacuum expectation value, there is no $Z$--$Z'$ mixing at this
stage.} of the $Z'$. The mass-squared matrix for the two scalars
$\xi_{1,2}$ reads
\[
    M^2_{\xi_i} = \left( \begin{array}{ccc} 
                 \lambda_1 x_1^2 & \quad & \lambda_{12} x_1 x_2 \\[1ex]
                 \lambda_{12} x_1 x_2  & \quad & \lambda_1 x_1^2
                 \end{array} \right),
\]
leading to mass eigenstates $\tilde\xi_{1,2}$ defined by 
\begin{equation}
\left(\begin{array}{c} H_1 \\[1ex] H_2 \end{array}
\right) 
= 
\left(\begin{array}{ccc} \cos \alpha_\chi & \quad & \sin\alpha_\chi
             \\[1ex] -\sin \alpha_\chi & \quad & \cos\alpha_\chi
 \end{array}
\right) \;
\left(\begin{array}{c} \xi_1 \\[1ex] \xi_2 \end{array} \right) 
\ , \qquad \tan (2\alpha_\chi) = 
     \frac{2 \lambda_{12} x_1 x_2}{\lambda_1 x_1^2 - \lambda_2 x_2^2},
\end{equation}
with the corresponding masses being
\begin{equation}
M^2_{H_1,H_2} = \frac{1}{2} \left[ \lambda_1 x_1^2 + \lambda_2 x_2^2
             \pm |\lambda_1 x_1^2 - \lambda_2 x_2^2| \;
     {\rm sec}(2 \alpha_\chi) \right].
\end{equation}

%\normalsize
%
%\vskip 10pt
%\hrule
%\vskip 10pt

\subsection{Fermionic sector and anomalies}

Since one of our primary goals is to explain neutrino masses and
mixings, we must include extra neutral fermions and at least two of
them. This is the only addition we propose in this sector, and as we
would shortly see, invoking three such right-handed fields is not only
enough to ensure the cancellation of all possible
anomalies\footnote{While, gauge anomalies can also be canceled using
    the Green-Schwarz mechanism---we refer the reader to
    Ref.~\cite{Ekstedt:2017tbo} for a phenomenological discussion 
    of such constructions---we eschew this in favour of a more 
    canonical approach.}, but also
leads to very interesting phenomenological consequences.  

For the sake of simplicity, we consider the $\mathrm{U}(1)_z$ charges to be
family-blind, as far as the SM fermions are concerned, denoting these
to be $z_Q$ (quark doublets), $z_L$ (lepton doublets), $z_u, z_d$ (the
right-handed up-like and down-like quarks respectively) and $z_e$ for
the right-handed charged leptons. Similarly, three RHN fields $N_i$ are assigned charges $z_i$, not necessarily
equal. Before we consider the fermion masses and, thereby, relate these 
to $z_H$ and $z_{\chi_A}$, let us first discuss the anomalies defined 
as ${\cal A} \equiv {\rm tr}_L(T_a T_b T_c) - {\rm tr}_R(T_a T_b T_c)$ 
where $T_a$ are the symmetry generators and the traces are over left- and
right-handed fermions. 
The SM gauge anomalies, of course, remain unaltered and the only nontrivial 
quantities are those pertaining to $\mathrm{U}(1)_z$ and are listed below
\begin{center}
\begin{tabular}{|c|c|}
\hline
Anomaly & Expression \\
\hline
$\left[\mathrm{SU}(3)_c\right]^2 \mathrm{U}(1)_z$ & $2 z_Q = z_u + z_d $
\\[1ex]
$\left[\mathrm{SU}(2)\right]^2 \mathrm{U}(1)_z$ & $3 z_Q + z_L = 0$
\\[1ex]
$\left[\mathrm{U}(1)_Y\right]^2 \mathrm{U}(1)_z$ & $z_Q + 3 z_L = 8 z_u + 2 z_d + 6 z_e$
\\[1ex]
$\mathrm{U}(1)_Y \left[\mathrm{U}(1)_z\right]^2$ & $z_Q^2 - z_ = 2 z_u^2 - z_d^2 - z_e^2$
\\[1ex]
$\left[\mathrm{U}(1)_z\right]^3$ & $\displaystyle
 6 z_Q^3 + 2 z_L^3 = 3 z_u^3 + 3 z_d^3 + z_e^3+ \sum_{i=1}^3 z_i^3 $
\\[1ex]
\hline
\end{tabular}
\end{center}

It is easy to see that, using the first two conditions, the third simplifies 
to $ 2 z_Q + z_u + z_e = 0$. Similarly, the first three, together, imply
that the fourth one is satisfied identically. And, finally, the fifth one
simplifies to $\sum_{i=1}^3 z_i^3 = 3 \, (z_u - 4 z_Q)^3$. It is also easy to see
that the mixed gauge-gravity anomaly ($R^2 \mathrm{U}(1)_z$) does not present 
an independent constraint. 

\subsection{$\mathbf{U(1)_z}$ charge assignment}
The existence of mass terms for the charged fermions demands that 
\begin{equation}
z_H = z_L - z_e = z_Q - z_d = z_u - z_Q \ .
\end{equation}
Note that only one of these equations is independent once anomaly
cancellations have been imposed (in fact, just the
$\left[\mathrm{SU}(3)_c\right]^2 \mathrm{U}(1)_z$ and $\left[\mathrm{SU}(2)\right]^2 \mathrm{U}(1)_z$ are
enough).  The $\mathrm{U}(1)_z$ charges of the SM fields can, then, be
expressed in terms of just two parameters, say $z_u$ and $z_Q$. Note,
however, that, for any $\mathrm{U}(1)$ theory, one combination of charges can
always be taken to be unity, without any loss of generality.  In the
present case, we shall choose $z_u - 4 z_Q = 1$ and consider $z_Q$ to
be the remaining free parameter. The consequent charge assignments have been 
displayed in Table.\ref{tab:charges}.

%\vskip 10pt
%\hrule
%{\em This bit should go elsewhere. However, I am writing it here so that 
%one remembers about it.}
%
%One of the strongest constraints on such a scenario would emanate from
%$Z'$ production at the LHC with a subsequent decay into a
%low-background final state, such as a lepton-pair. It is amusing to
%note that choosing $z_Q$, hitherto a free parameter, appropriately
%would minimize the production cross-section without affecting the rest
%of the phenomenology. For example, to the leading order, 
%\[
%\sigma(pp \to Z' + \mbox{rest}) \propto (z_q^2 + z_u^2) \, {\cal F}_u 
%                                    + (z_q^2 + z_d^2) \, {\cal F}_d \ .
%\]
%Here, the flux ${\cal F}_{u}$ is given by 
%\[
%{\cal F}_u \equiv \int_{m_{Z'}^2/s}^1 \frac{dx}{x} \, 
%            \left[f_u(x, Q^2) f_{\bar u}\left(\frac{m_{Z'}^2}{s \, x}, Q^2\right)
%                 + f_{\bar u}(x, Q^2) f_{}\left(\frac{m_{Z'}^2}{s \, x}, Q^2\right)
%                 \right]
%\]
%with $\sqrt{s}$ being the total center-of-mass energy available at the 
%LHC, and $f_j(x, Q^2)$ the density of the $j^{\rm th}$ parton for a given 
%momentum fraction $x$ and computed at the scale $Q$. An analogous expression 
%holds for  ${\cal F}_{d}$. While the ratio of the two fluxes (those for 
%the other quark flavours being subdominant) is a function of $m_{Z'}^2$, it 
%is intriguing to note that $\sigma(pp \to Z' + \mbox{rest})$ is minimized 
%for $z_Q \sim -1/4$. 
%
%
%\vskip 10pt
%\hrule
%\vskip 10pt

The charges $z_i$ for the $N_{iR}$ fields, thus, need to satisfy 
\[
   \sum_i z_i^3 = 3 \ ,
\]
and the solution space is a two-dimensional one. Restricting
ourselves to rational values, the simplest assignment would be $z_i =
1$, a choice that has been explored in a different
context~\cite{Appelquist:2002mw}. This, though, is unsuitable as far 
as neutrino mass generation is concerned. Consequently, we adopt
the next simplest choice, namely $z_{1,2} = 4, \ z_3 = -5$.

\begin{table}
\centering
\vspace{0.5em}
\begin{tabular}{|ccccl|}
\hline
&$\mathrm{SU}(3)_{c}$&$\mathrm{SU}(2)_{L}$&$\mathrm{U}(1)_{Y}$&$\mathrm{U}(1)_{X}$\\
\hline
$q_{L}$&3&2&$1/6$&$z_{Q}$\\
$u_{R}$&3&1&$2/3$&$1 + 4 z_{Q}$\\
$d_{R}$&3&1&$-1/3$&$-1 - 2z_{Q}$\\
$\ell_{L}$&1&2&$-1/2$&$-3z_{Q}$\\
$e_{R}$&1&1&$-1$&$-1 -6z_{Q}$\\
$H$&1&2&$1/2$&$1 + 3z_{Q}$\\
\hline
$N_{1R}, N_{2R}$&1&1&0&$4$\\
$N_{3R}$&1&1&0&$-5$\\
$\chi_1$&1&1&0&$z_{\chi_1}$\\
$\chi_2$&1&1&0&$z_{\chi_2}$\\
\hline
\end{tabular}
\caption{The charge assignments for the fermions and scalars of the model.}\label{tab:charges}
\end{table}

\section{Neutrino masses}
\label{sec:numass}
While the Yukawa (and, hence, the mass) terms for the charged fermions
proceed as within the SM, {\em viz.},
\[
  {\cal L}_{\rm Yuk.} = y^{u}_{ij} \bar Q_{Li} u_{Rj} \widetilde H 
                    + y^{d}_{ij} \bar Q_{Li} d_{Rj} H 
                    + y^{e}_{ij} \bar L_{Li} e_{Rj} H + \textrm{H.c.} \ ,
\]
where $\widetilde H = i \sigma_2 H^*$, dimension-4 gauge invariant
Yukawa (or even bare mass) terms are not possible for the
neutrinos. The situation changes if
the theory is treated not as a fundamental one, but only as the
  low-energy limit of some more fundamental theory operative at some
  scale $\Lambda$ or higher. Freed of the restriction of being
  renormalizable, the effective field theory
would admit higher-dimensional terms of the form\footnote{While analogous
    terms can be written for the charged fermions as well, these would
    be subdominant to the usual Yukawa terms and we omit all
    discussions thereof.}
\begin{equation}
\begin{array}{rcl}
{\cal L}_{\nu \rm mass} &= & \displaystyle {\cal L}_{\rm Dirac} + {\cal L}_{\rm Wein.};
\\[2ex]
{\cal L}_{\rm Dirac} & = & \displaystyle \sum_{i= 1}^3 \sum_{\alpha = 1}^2 
        y_{i \alpha} \bar L_{iL} N_{\alpha R} \widetilde H \; 
            \frac{\chi_1^{a_1} \, \chi_2^{a_2}}{\Lambda^{|a_1|+ |a_2|}}
          + \sum_{i= 1}^3  \tilde y_{i} \bar L_{iL} N_{3 R} \widetilde H  \;
            \frac{\chi_1^{a_3} \, \chi_2^{a_4}}{\Lambda^{|a_3|+ |a_4|}} + \textrm{H.c.};
\\[2ex]
{\cal L}_{\rm Wein.} & = & \displaystyle \sum_{i,j= 1}^3 
        w_{i j} \overline{L^c_{iL}} L_{jL} H H 
            \frac{\chi_1^{b_1} \, \chi_2^{b_2}}{\Lambda^{|b_1|+ |b_2|+1}}
        + \sum_{\alpha,\beta= 1}^2  s_{\alpha\beta} \overline{N^c_{\alpha R}} N_{\beta R} 
            \frac{\chi_1^{b_3} \, \chi_2^{b_4}}{\Lambda^{|b_3|+ |b_4|-1}}
\\[2ex]
& + & \displaystyle
    \sum_{\alpha=1}^2  s_{\alpha3} \overline{N^c_{\alpha R}} N_{3 R} 
            \frac{\chi_1^{b_5} \, \chi_2^{b_6}}{\Lambda^{|b_5|+ |b_6|-1}}
        +  s_{33} \overline{N^c_{3R}} N_{3 R} 
            \frac{\chi_1^{b_7} \, \chi_2^{b_8}}{\Lambda^{|b_7|+ |b_8|-1}} + \textrm{H.c.},
\end{array}
\end{equation}
where the couplings $y_{i\alpha}, \tilde y_i, w_{ij}, s_{\alpha\beta},
s_{\alpha3}$ and $s_{33}$ are dimensionless and the exponents satisfy
\begin{equation}
\begin{array}{rcl c rcl}
z_{\chi_1} a_1 + z_{\chi_2} a_2 & = & -3  & \qquad \quad &
z_{\chi_1} a_3 + z_{\chi_2} a_4 & = & 6
\\[1ex]
z_{\chi_1} b_1 + z_{\chi_2} b_2 & = & -2  & \qquad \quad &
z_{\chi_1} b_3 + z_{\chi_2} b_4 & = & -8
\\[1ex]
z_{\chi_1} b_5 + z_{\chi_2} b_6 & = & 1  & \qquad \quad &
z_{\chi_1} b_7 + z_{\chi_2} b_8 & = & 10 \ .
\end{array}
\end{equation}
It should be realized that only integer solutions for the exponents
are permissible as non-integral values would imply nonlocal
operators. Negative values for the exponents are to be interpreted as
positive powers of $\chi_1^* \, (\chi_2^*)$ as the case may be.

Before discussing the ramifications of ${\cal L}_{\nu {\rm mass}}$, it
is instructive to remind ourselves of the possible origin of the
same. As can be readily recognized, these can arise from a UV-complete
theory once a slew of fields (especially fermionic ones) are
integrated out. Clearly, these fields
must have masses larger than $\Lambda$ and carry nonzero $\mathrm{U}(1)$
charges. It might be argued, then, that the requirement of the effective
theory being anomaly-free is a superfluous one, for the anomaly(ies),
being a child of the UV regularization, could, in principle, be
canceled by the heavy fermions $\Psi_i$. However, note that the
$\Psi_i$ themselves should be vector-like, as else their masses can
only arise from spontaneous breaking of the $\mathrm{U}(1)$ symmetry and, thus,
should be below $\Lambda$. And since vector-like fermions do not
contribute to gauge anomalies, the effective theory better be
anomaly-free.

A further issue pertains to the relative strengths of tree-order and
loop-level contributions in the generalization of the FN-mechanism
that our theory really represents. Consider a typical term in ${\cal
  L}_{\nu {\rm mass}}$ which has $n$ powers of, say, $\chi_1$. Letting
the $\chi_1$ lines go into the vacuum (courtesy spontaneous symmetry
breaking) gives us a factor of $(x_1/\Lambda)^n$. On the other hand,
closing a $\chi_1$ loop would, typically, give us a factor of ${\cal
  O}(\Lambda^2 / 16 \pi^2 x_1^2)$, as the loop momentum would need to
be cut off at the scale $\Lambda$.  Thus, the exclusion of loops is
well motivated for $\Lambda > x_j \gapp \Lambda/(4\pi)$, inequalities that we
would satisfy in further calculations.

With each of the terms in ${\cal L}_{\rm Wein.}$ violating
lepton-number, it is tempting to characterize the corresponding mass
terms, realised on breaking the $\mathrm{U}(1)_z$
symmetry, as Majorana masses. An argument against this would be the
fact that, in the canonical sense, a Majorana particle may not have
any nonzero additive quantum numbers, whereas each of $\nu_{iL}$ and
$N_{iR}$ certainly do. Rather, these terms should be thought of as the
generalization of the Weinberg-operator~\cite{Weinberg:1979sa} that is
allowed as a dimension-5 correction to the SM.  Indeed, on the
breaking of the $\mathrm{U}(1)_z$ symmetry, the
corresponding charge is no longer a valid quantum number in the
ensuing theory, and the mass term generated thereupon can indeed be
thought of as a Majorana mass.

As a very specific case, let us consider the assignment 
\beq
    z_{\chi_1} = -3/4 \ , \qquad z_{\chi_2} = -4 ,
\eeq
which leads to rather interesting phenomenology\footnote{It should be realised 
that this choice is not a special one and qualitatively similar results
would be obtained for many other choices.}. With this choice, 
the masses for $N_{1,2}$ sub-sector are relatively unsuppressed, while terms
connecting $N_{3R}$ are highly suppressed. Indeed, retaining just the least 
suppressed terms in each sector would lead to

\beq
\barr{rcl}
{\cal L}_{\rm Dirac} & = & \displaystyle \sum_{i= 1}^3 \sum_{\alpha = 1}^2 
        y_{i \alpha} \bar L_{iL} N_{\alpha R} \widetilde H \; 
            \frac{\chi_1^{4} }{\Lambda^{4}}
          + \sum_{i= 1}^3  \tilde y_{i} \bar L_{iL} N_{3 R} \widetilde H  \;
            \frac{\chi_1^{*8} }{\Lambda^{8}} + \textrm{H.c.}\ ,
\\[2ex]
{\cal L}_{\rm Wein.} & = & \displaystyle \sum_{i,j= 1}^3 
        w_{i j} \overline{L^c_{iL}} L_{jL} H H 
            \frac{\chi_1^{8} \, \chi_2^{*}}{\Lambda^{10}}
        + \sum_{\alpha,\beta= 1}^2  s_{\alpha\beta} \overline{N^c_{\alpha R}} N_{\beta R} 
            \frac{\chi_2^{2}}{\Lambda}
\\[2ex]
& + & \displaystyle
    \sum_{\alpha=1}^2  s_{\alpha3} \overline{N^c_{\alpha R}} N_{3 R} 
            \frac{\chi_1^{4} \, \chi_2^{*}}{\Lambda^{4}}
        +  s_{33} \overline{N^c_{3R}} N_{3 R} 
            \frac{\chi_1^{8} \, \chi_2^{*4}}{\Lambda^{11}} + \textrm{H.c.}\ .
\end{array}
 \label{nu_mass_our}
 \eeq
The terms corresponding to $w_{ij}$ and $s_{33}$ are too 
small to be of any consequence, and, formally, could be dropped altogether 
if we restrict ourselves to operators of mass dimension 12 or less. 
 This leads to 
\beq
\barr{rcl}
{\cal L}_{\nu {\rm mass}}  &\approx & \dis {\cal L}^{(5)} + {\cal L}^{(8)} 
                                    + {\cal L}^{(12)} + \textrm{H.c.}
\\[2ex]
\dis {\cal L}^{(5)} & \equiv & \dis 
        \sum_{\alpha,\beta= 1}^2  s_{\alpha\beta} \overline{N^c_{\alpha R}} N_{\beta R} 
            \frac{x_2^{2}}{\Lambda},
\\[2ex]
\dis {\cal L}^{(8)} & \equiv & 
\dis \sum_{i= 1}^3 \sum_{\alpha = 1}^2 
        y_{i \alpha} \bar L_{iL} N_{\alpha R} \widetilde H \; 
            \frac{x_1^{4} }{\Lambda^{4}}
    + \sum_{\alpha=1}^2  s_{\alpha3} \overline{N^c_{\alpha R}} N_{3 R} 
            \frac{x_1^{4} \, x_2^{*}}{\Lambda^{4}},
\\[2ex]
\dis {\cal L}^{(12)} & \equiv & \dis 
\sum_{i= 1}^3  \tilde y_{i} \bar L_{iL} N_{3 R} \widetilde H  \;
            \frac{x_1^{*8} }{\Lambda^{8}} \ ,
\earr
\label{nu_mass_simpl}
\eeq
and this is the form that we would be working with henceforth.

\subsection{Identifying the mass eigenstates}
The neutrino mass matrix, as given in Eq.~\eqref{nu_mass_simpl} can be
represented, in the $(\nu_j, N_3, N_1, N_2)$ basis, by \beq {\cal
  M}_\nu = \left( \barr{cc} 0_{3\times 3} & {\cal D} \\ D^T & M_N
\earr \right) .  \eeq Denoting $\xi \equiv x/\Lambda$, where $x$
\hldc{is} either of $x_{1,2}$ (we assume that there is no large
hierarchy between the $x_i$), the matrices above have structures
%%%%%%
\beq 
{\cal D} \approx v \xi^4 \left(
\begin{array}{ccc}
 y_{13} \xi^4 & y_{11} & y_{12} \\
 y_{23} \xi^4 & y_{21} & y_{22} \\
 y_{33} \xi^4 & y_{31} & y_{32} \\
\end{array}
\right)
 \qquad
M_N \sim \frac{x^2}{\Lambda} \, \left(
\begin{array}{ccc}
0 & s_{31} \xi^3 & s_{32} \xi^3  \\
s_{31} \xi^3 & a_1 & 0 \\
s_{32} \xi^3 & 0 & a_2 \\
\end{array}
\right).
\eeq
Note that the $N_1$--$N_2$ sub-sector
of the mass matrix can be chosen to be diagonal without any loss of
generality, with $a_1$ and $a_2$  as the coefficients.

While two eigenvalues of $M_N$ are large, {\em viz.} ${\cal
O}(x^2/\Lambda)$, these heavy neutrinos still tend to be lighter than
the $Z'$, owing to the smallness of $\xi$ (in comparison to $g_X$).
The third eigenvalue of $M_N$ is much smaller, namely only $\lambda_3 \sim {\cal
O}(\xi^6 x^2/\Lambda)$. This state has only a small mixing with the
heavier ones, with the mixing angles being ${\cal O}(\xi^4)$.

More importantly, it might seem that a straightforward application of
the seesaw mechanism may not be possible, given that one of the
eigenvalues of $M_N$ is smaller than some of the Dirac masses.
However, note that the elements ${\cal D}_{i1}$ are smaller than the
corresponding fulcrum $\lambda_3$ by at least a factor of
$v\xi/x$. Consequently, the seesaw mechanism goes through trivially,
and even the first-order estimate is rather accurate.  This is
corroborated by a full numerical calculation of the eigensystem of the 
full $6\times6$ mass matrix.

On block-diagonalization, we have the effective mass matrix for the
light-sector to be given by
\beq
M_{3\times 3}  =  - {\cal D} M_N^{-1} {\cal D}^T  + {\cal O}(M_N^{-2}).
\eeq
Explicitly,
%%%%
\beq
M_{3 \times 3} 
= \frac{-v^2 \xi^6}{2 \Lambda} \left(
\begin{array}{ccc}
M_{11} 
& 
M_{12} 
& M_{13} \\
M_{12} 
& M_{22} 
& M_{23} \\
M_{13} & M_{23} & M_{33} \\
\end{array}
\right)\ .
  \label{block_diag_1}
\eeq
%%%%%
where,
%%%%%
\beq
\barr{rcl}
M_{11} &=& \dis  ({y_{11}}-{y_{12}})^2+2 \xi ({y_{11}}+{y_{12}}) {y_{13}}-
\xi^2 y_{13}^2 \\[1ex]
M_{12} &= & \dis 
({y_{11}}-{y_{12}}) ({y_{21}}-{y_{22}})-\xi^2 {y_{13}} {y_{23}}+\xi ({y_{13}} ({y_{21}}+{y_{22}})+({y_{11}}+{y_{12}}) {y_{23}})
\\[1ex]
M_{13} &= & \dis  ({y_{11}}-{y_{12}}) ({y_{31}}-{y_{32}})-\xi^2 {y_{13}} {y_{33}}+\xi ({y_{13}} ({y_{31}}+{y_{32}})+({y_{11}}+{y_{12}}){y_{33}})
\\[1ex]
M_{22} &= & \dis  ({y_{21}}-{y_{22}})^2+2 \xi ({y_{21}}+{y_{22}}) {y_{23}}-\xi^2 y_{23}^2 
\\[1ex]
M_{23} &= & \dis  ({y_{21}}-{y_{22}}) ({y_{31}}-{y_{32}})-\xi^2 {y_{23}} {y_{33}}+\xi ({y_{23}} ({y_{31}}+{y_{32}})+({y_{21}}+{y_{22}})
{y_{33}})
\\[1ex]
M_{33} &= & \dis  ({y_{31}}-{y_{32}})^2+2 \xi ({y_{31}}+{y_{32}}) {y_{33}}-\xi^2 y_{33}^2
\earr
  \label{block_diag_2}
\eeq

Several points demand attention:
\begin{itemize}
\item As argued above,  dropping the higher order terms in 
      in $M_{3\times 3}$ is an excellent approximation. Similarly, the 
      mixing between the $\nu_i$ and $N_{\alpha}$ is quite small and the 
      eigenstates of $M_{3\times 3}$ are predominantly doublets. 

\item If the Yukawa couplings are real 
({\em i.e.}, if there is no CP violation in this sector), then the
 real symmetric matrix $M_{3\times3}$ is given in terms of 6
 parameters.  These can be uniquely determined in terms of the three
 light neutrino masses and the three mixing angles. 

\item {\em A priori}, the matrix ${\cal D}$ is defined by nine  
      parameters (assumed to be real), and further parameters appear 
      in $M_N$. However, only six independent combinations may enter 
      $M_{3\times3}$, with the rest serving only to define the heavy 
      sector and the tiny mixing between the heavy and light sectors.

\item Finally, while exploring the parameter space,
it is important to ensure that $(M_{3\times3})_{ee}$ is bound by the non-observation of 
neutrinoless double beta decay~\cite{Das:2018ldy}.
\end{itemize}

%\subsection{The current experimental data}
%Experimentally,

%\subsection{Neutrino Plots}
%%%%%% Deleted part %%%%%%%%%%%%%%%%
%Now, each column of $U$ is a normalized eigenvector of $M_{3\times3}$.
%Identifying the third column with the massless state $|\nu_3\rangle$, 
%we have
%\beq
%\dis \tan\theta_{23} = \frac{d_{12} d_{31}-d_{11} d_{32}}{d_{12} d_{21}-d_{11} d_{22}} 
%\ .
%\eeq
%Similarly, 
%\beq
%\frac{d_{21} d_{32}-d_{22} d_{31}}{d_{12} d_{21}-d_{11} d_{22}}
%= \tan\theta_{13} \, \sqrt{1 + \tan^2\theta_{23}} \ ,
%\eeq
%and
%\beq
%c^2_{12} = (\nu_1 - \nu_2)^{-1} \, 
%\left[ \left(\mu_1 d_{11}^2+ \mu_2d_{12}^2 \right) (1 + \tan^2\theta_{13}) 
%   - \nu_2 \right] \ .
%\eeq
%These three  relations, alongwith the relations for the 
%masses can be used to fix all six relevant parameters.

\subsection{Neutrino Results}
Since the heavy-light mixing is tiny, low-energy experiments are
well-described in terms of the light neutrinos alone. To this end, we diagonalize the light part of the mass matrix through
%%%%
\beq
U^T M_{3\times 3} U = {\rm diag}(m_1, m_2, m_3),
\eeq
where the PMNS matrix $U$ is given by
\[
U^T = R_{12}(\theta_{12}) \,  R_{13}(\theta_{13}) \, R_{23}(\theta_{23}),
\]
with $R_{jk}$ denoting a rotation in the $jk$ plane through an angle $\theta_{jk}$.
In other words,
\beq
U = \left(
\begin{array}{ccc}
 c_{12} c_{13} & c_{13} s_{12} & s_{13} \\ -s_{12} c_{23}-c_{12} s_{13} s_{23} & c_{12} c_{23}-s_{12} s_{13} s_{23} & c_{13} s_{23} \\
 s_{12} s_{23}-c_{12} c_{23} s_{13} & -c_{12} s_{23}-c_{23} s_{12} s_{13} & c_{13} c_{23} \\
\end{array}
\right),
\eeq
where $c_{ij} \equiv \cos\theta_{ij}$ and $s_{ij} \equiv \sin\theta_{ij}$.

%\section{Results}
The different Yukawa parameters are varied such that the mass squared
differences of neutrino mass eigenstates namely $m_i, i=1,2,3 $
defined as, $$ \Delta_{21} = m_2^2 - m_1^2 $$ and $$ \Delta_{32} = \pm (m_3^2 - m_2^2) $$ along with three mixing angles
$\theta_{12}, \theta_{13}, \theta_{23}$ conform to the
experimental results as listed in Table.~\ref{Nucons}. Another
important constraint which is to be satisfied is 
 cosmological one on the sum of the masses of the light stable neutrinos, namely,
\cite{Capozzi:2017ipn}:
\begin{equation}
\sum_i m_i < (0.340-0.715)~\textrm{eV}\ , 
\label{Nutm}
\end{equation}
with 95\% confidence level (CL). A more stringent limit can be obtained if we take into account Baryonic Acoustic Oscillations~\cite{Ade:2015xua,Vagnozzi:2017ovm}. 
But we have stuck to the limit above.

\begin{table}
\begin{center}
\begin{tabular}{ |c|c|c|c|c|c| } 
 \hline
 & $\Delta_{21} (\rm eV^2)$ & $\Delta_{32} (\rm eV^2)$ & $ \sin^2\theta_{23} $ & $ \sin^2\theta_{13} $ & $ \sin^2\theta_{12} $ \\ 
 \hline
 NH   & $7.39_{-0.20}^{+0.21} \times 10^{-5} $  & $2.449_{-0.030}^{+0.032} \times 10^{-3} $  &  $0.558_{-0.33}^{+0.20}$ &  $2.241_{-0.65}^{+0.66} \times 10^{-2} $ & $0.31_{-0.12}^{+0.13} $ \\ 
 \hline
 IH  & $7.39_{-0.20}^{+0.21} \times 10^{-5} $ & $-2.509_{-0.032}^{+0.032} \times 10^{-3} $ &  $0.563_{-0.26}^{+0.19} $ & $2.261_{-0.64}^{+0.67} \times 10^{-2} $  & $0.31_{-0.12}^{+0.13} $ \\ 
 \hline
 \end{tabular}
\caption{Bounds on neutrino mass and mixing parameters from neutrino
  oscillation experiments, for both normal hierarchy (NH) and inverted
  hierarchy (IH) of the neutrino mass spectrum. The error bars shown 
    correspond to $1\sigma$~\cite{Esteban:2018azc}.}
\label{Nucons}
\end{center}
\end{table}

Clearly, we have more parameters in the theory than there are 
data in the neutrino sector. Consequently, some parameters are indeterminable,
and must be fixed by hand. To ease the task of identifying the crucial 
dependence, we choose to make some simplifying assumptions:
\begin{itemize}
\item While we had already assumed that there was no large 
  hierarchy between the vacuum expectation values for the two new scalars,
  we eliminate any choice and consider only $x_1=x_2 = 10$ TeV.
\item The cutoff scale $\Lambda$ is fixed at $100$ TeV; in other words,
   $\xi = 0.10$.
\item The two heavy RHNs, predominantly $N_{1,2}$, 
  should have a mass ${\cal O}(\xi x)$ and these are 
  held at $1.2$~TeV and $1.25$~TeV respectively. The splitting between them 
  is not germane to the discussion at hand, and has been incorporated 
  just to ensure that the numerical results are not affected by degeneracy.
\item The third RHN ($N_3$-like) is held at $6$~keV. 
  This can be achieved, for example, if $s_{\alpha 3}
\approx 0.05$. in Eq.~\eqref{nu_mass_our}.
As it turns out, certain results are quite dependent on this (although, 
  not the light neutrino phenomenology) and we shall return to this in a later 
  section.
\end{itemize}

  It should be realized that the above choices are not special in
  any way and do not leave qualitative impact on the determination of
  the rest of the parameter space. These are to be fixed by an
  analysis of the neutrino oscillation results. Clearly, this last bit
  would be dependent on the two different neutrino mass hierarchies
  that are experimentally viable, and we consider them in turn. In
  doing this, it needs to be borne in mind that, for a very large part
  of the parameter space, one of the three SM-like neutrinos is
  distinctly lighter than the others. This is but a consequence of the
  fact that, were $N_3$ to be decoupled, the light mass matrix
  $M_{3\times 3}$ (resulting from the seesaw mechanism) would be a
  rank-2 one.

%\newpage

\subsubsection{Normal Hierarchy}
%\subsubsection{Normal Hierarchy}
Terming the lightest mass $m_1 \sim 0$, the two other mass eigenvalues 
would be $m_2 \sim \sqrt{\Delta_{21}}$ and $m_3 \sim \sqrt{\Delta_{32}+\Delta_{21}}$, thereby
satisfying the hierarchy $m_1<m_2 \ll m_3$.

%\sout{In this case, all the allowed regions of different Yukawa coupling parameter space consists of two distinct areas which get more prominent with the imposition of the constraints like $\Delta m_{21}^2, \Delta m_{32}^2$. }
%
\begin{figure}
\begin{center}
\includegraphics[scale=.5]{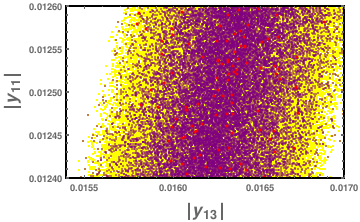} \hspace{0.2cm}
\includegraphics[scale=.5]{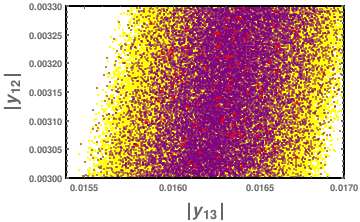}
\includegraphics[scale=.5]{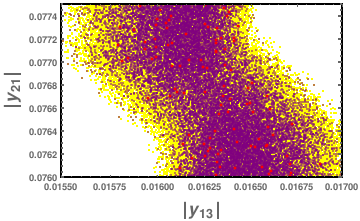}\hspace{0.2cm}
\includegraphics[scale=.5]{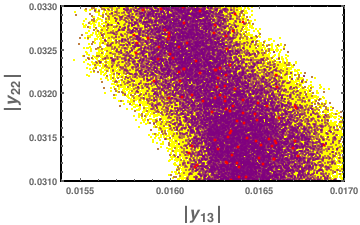}
\includegraphics[scale=.7]{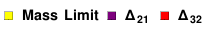}
\caption{\label{nh1} Correlation of Yukawa couplings in the Dirac sector for
  neutrino masses in normal hierarchy. Allowed points after
  diagonalization of neutrino mass matrix satisfying the bound on
  total mass of three neutrino species (in yellow), points with
  satisfying the bound on $\Delta m_{32}^2$ (in purple) and allowed
  points after another bound of $\Delta m_{12}^2$ (in red).}
  \end{center}
\end{figure}

  As is well known, the requirement of $\theta_{23} \sim 45^\circ$
  imposes strong constraints on the neutrino mass matrix, and is often
  sought to be explained by family symmetries. In the present context,
  this is to be ensured by judiciously choosing the couplings $y_{ij}$
  guided by Eq.~\eqref{block_diag_2}. While many different solutions
  are possible, given that $\xi$ is small, if we want to eschew very
  large hierarchies in the couplings and/or large cancellations, we
  should look for the possibility that $y_{21}- y_{22} \approx y_{31}
  - y_{32} \sim 0.01$. The first approximate equality ensures that the
  leading contributions are of the same order, while the second one
  ensures that cancellations are not extreme (since the couplings
  themselves would turn out to be of a similar size).  The correct value for the heaviest
neutrino mass $m_3$ is obtained for a somewhat larger value of
  $y_{33}$, namely $y_{33} \sim {\cal O}(0.1)$. As
  Fig.~\ref{nh1} shows, it is possible to satisfy all the neutrino
  constraints in our framework with these Yukawas, without resorting
  to very small values as is often the case for general seesaw models. Indeed, relatively larger values of the Yukawas $y_{21} \sim 0.08, y_{22}
  \sim 0.03$ as shown in Fig.~\ref{nh1} play a pivotal role to have a
  significant mixing in the $1-2$ sector. It should be realized
    that Fig.~\ref{nh1} does not reflect the entire viable parameter
    space.  The very structure of Eq.~\eqref{block_diag_1}, for
    example, stipulates that the interchange $(y_{11}, y_{21})
    \leftrightarrow (y_{12}, y_{22})$ would result in identical masses
    and mixings. Rather, the parameter space displayed in
    Fig.~\ref{nh1} should be considered a representative set of
    solutions.

\begin{figure}
\begin{center}
\includegraphics[scale=.5]{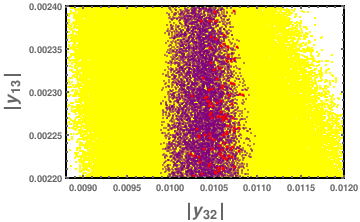} \hspace{0.2cm}
\includegraphics[scale=.5]{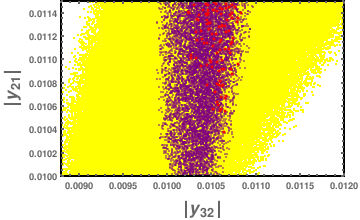}
\includegraphics[scale=.5]{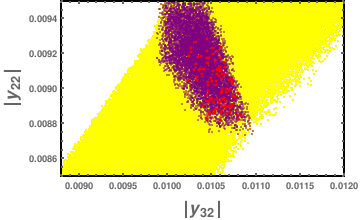} \hspace{0.2cm}
\includegraphics[scale=.5]{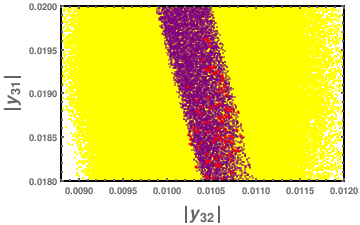}
\includegraphics[scale=.7]{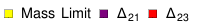}
\caption{\label{ih1} Correlation of Yukawa couplings in the Dirac sector for an inverted hierarchy of neutrino masses.}
\end{center}
\end{figure}

\subsubsection{Inverted Hierarchy}
%\subsubsection{Inverted Hierarchy}
Similar to the normal hierarchy, we have taken a simplified scenario where
$m_3$ is now taken to be zero. Other two mass eigenvalues in this case then become $m_1 \sim \sqrt{\Delta_{23}-\Delta_{21}}$ and $m_2 \sim \sqrt{\Delta_{23}}$. This satisfies the hierarchy $m_3 \ll m_1 \approx m_2$.

Similar in the previous case, some initial assumption on Yukawa couplings are taken such as $y_{11} = y_{21} + 0.01$ and $y_{12} = y_{22} + 0.01$, such that we have similar contribution from ($y_{21}- y_{22}$) and ($y_{11} - y_{12}$). This is required to satisfy the mass of $m_1$ to be non-zero along with the significant mixing with $\theta_{12} \sim 33^o$, in the $1-2$ sector. There are two more distinct regions of Yukawas with magnitude of similar order to those shown in the plots where all the neutrino constraints are satisfied. In this case also, the allowed Yukawas values satisfying the neutrino results are significant enough, as presented in the Fig.~\ref{ih1}. In the allowed parameter space, the difference $y_{31} - y_{32}$ is relatively larger than the difference $y_{21} -y_{22}$, as depicted in the Fig.~\ref{ih1}, helps to have a negligibly small $m_3$ term along with significant mixing in the form of $\theta_{23} \sim 45^o$.

%\sout{A major portion of parameter space is available that diagonalizes the neutrino mass matrix along with satisfying the constraint on the neutrino total mass. Unlike the normal hierarchy case, here the $M_{11}^{bd}$ term cannot be very small to arrange for non zero $m_1$ values after mixing through $U_{\rm PMNS}$ elements. That leads to larger Yukawa coupling as $y_{11} \sim 10^{-2}$, which after imposition of $\Delta_{12}^2, \Delta_{23}^2$ bounds restrict the $y_{11}$ to values smaller than $4.84 \times 10^{-2}$. It restricts the $y_{12} \le 0.03$, so that the difference remains relatively large i.e.  $y_{11}- y_{12} \sim 10^{-2}$. } 

%\sout{The mass difference $\delta m_{12}^2$ constrains the parameter space to allow Yukawa values with $|y_{22}| > 6 \times 10^{-3}$.}
%\newpage

\section{$Z^\prime$ phenomenology}
\label{sec:zppheno}

%In this section, we shall obtain the main constraints on the parameter space of our models. 

Until now, we have investigated the parameter space of the model only
in the context of the light neutrino masses and mixings.  As is
evident, these observables are sensitive primarily to the Yukawa
couplings (the Wilson coefficients in Eq.~\eqref{nu_mass_simpl}). While
there is a dependence on $x_i$ and $\Lambda$, these appear in a
trivial fashion, and as far as the limited number of observables
available at low energies are concerned, these dependencies can be
entirely subsumed in the Wilson coefficients\footnote{In other words,
  by trivially rescaling the Wilson coefficients, one can reproduce the observed
  masses and couplings for different sets of ($x_i, \Lambda$).}. And,
finally, this sector carries virtually no imprint of either the gauge
coupling $g_z$, nor the parameters in the Higgs potential.

The aforementioned parameters are best investigated at colliders,
either through direct production or by effecting precision
studies. Seven additional physical states now appear: a massive
neutral gauge boson $Z'$, two massive scalars $H_1$ and $H_2$, a
relatively light pseudoscalar $A$, two massive
(predominantly right-handed) neutrinos $N_{1,2}$ and, finally, a light
neutrino $N_3$ (again, predominantly right-handed). As for the
parameters in the gauge sector, apart from $g_z$ and $x_i$, we also
have the hitherto unfixed charge $z_Q$ (see Table~\ref{tab:charges}),
in terms of which the $\mathrm{U}(1)_z$ charges of all the SM fields were
specified, courtesy the requirements of anomaly cancellations.  That
$z_Q$ is still free is but a reflection of the fact that $\mathrm{U}(1)$
charges are, intrinsically, not quantized unless the symmetry had
descended from a bigger group.

With the $Z'$ having a substantial coupling to all the SM
fields\footnote{Note that while its coupling with any one set of the
  SM fermions can be switched off entirely, this cannot be done
  simultaneously for all of them.}, production of the $Z'$ at, say,
the LHC would be expected to constitute a sensitive probe of the
scenario. Interesting signatures of heavy neutrinos of our model can also be searched for at future lepton colliders~\cite{Banerjee:2015gca}. 
Before delving into the details of the $Z'$ phenomenology, it is amusing
to note that choosing $z_Q$, hitherto a free parameter, appropriately
would minimize the production cross-section without affecting the
neutrino phenomenology. For example, to the leading order,
%%%%
\beq
\sigma(pp \to Z' + X) \propto (z_q^2 + z_u^2) F_u 
                                    + (z_q^2 + z_d^2) F_d \ .
\label{cs_Zpr_prodn}
\eeq
%%%
Here, $X$ symbolizes the rest of the hadronic byproducts and the
flux $F_{u}$ is given by
\begin{equation}
F_u \equiv \int_{M_{Z'}^2/s}^1 \frac{dx}{x} \, 
            \left[f_u(x, Q^2) f_{\bar u}\left(\frac{M_{Z'}^2}{s \, x}, Q^2\right)
                 + f_{\bar u}(x, Q^2) f_{}\left(\frac{M_{Z'}^2}{s \, x}, Q^2\right)
                 \right],
\end{equation}
with $\sqrt{s}$ being the total center-of-mass energy available at the
LHC, and $f_j(x, Q^2)$ the density of the $j^{\rm th}$ parton for a
given momentum fraction $x$ and computed at the scale $Q$. An
analogous expression holds for $F_{d}$. While the ratio of the
two fluxes (those for the other quark flavours, being subdominant,
result in only a very minor correction) is a function of $M_{Z'}^2$,
it is intriguing to note that, for most of the region of interest,
$\sigma(pp \to Z' + X)$ is minimized for $z_Q \sim -1/4$. It is
interesting to note that a choice of $z_Q = -1/4$ is commensurate
(apropos $\mathrm{U}(1)_z$ charge quantization) with our assignments for the
$z_{\chi_i}$. Consequently, we use this value of $z_Q$ for much of the
numerical calculations below. However, this choice is not a very
crucial one and somewhat removed values of $z_Q$ might also be chosen
as illustrated by an alternative choice for some of the figures.

%\sout{There are, however, many free parameters present in the scalar and
%neutrino sectors. Those parameters will essentially determine the
%masses of the scalars and both left- and right-handed neutrinos.  In
%the $Z'$ phenomenology only the values of the masses will matter at
%the end. We, therefore, consider some benchmark values of these masses
%as input parameters in our analysis and do not explicitly show how
%they are related with the free parameters of the theory.} 
%
%\comment{This 
%point of view is untenable. The model exists in totality and is valid only 
%if it satisfies constraints from each sector, with the same set of 
%parameters. Else it remains questionable.}

\subsection{Branching ratios}

At a collider, of all the new states, the production of the $Z'$ is
the easiest.  Consequently, we begin by considering its branching
fractions. Owing to its large mass, the decay is dominated by the
two-body modes and these can be calculated trivially.  
We analytically compute various partial widths of two-body decay modes 
of the $Z'$ and the expressions thereof are given in Appendix~\ref{app:PW}. 
Our numerical results are cross-checked against
 \textsc{MadGraph}~\citep{Alwall:2014hca}. 

\begin{figure}
\centering
\subfloat[]{\includegraphics[height=5.5cm,width=5.2cm]{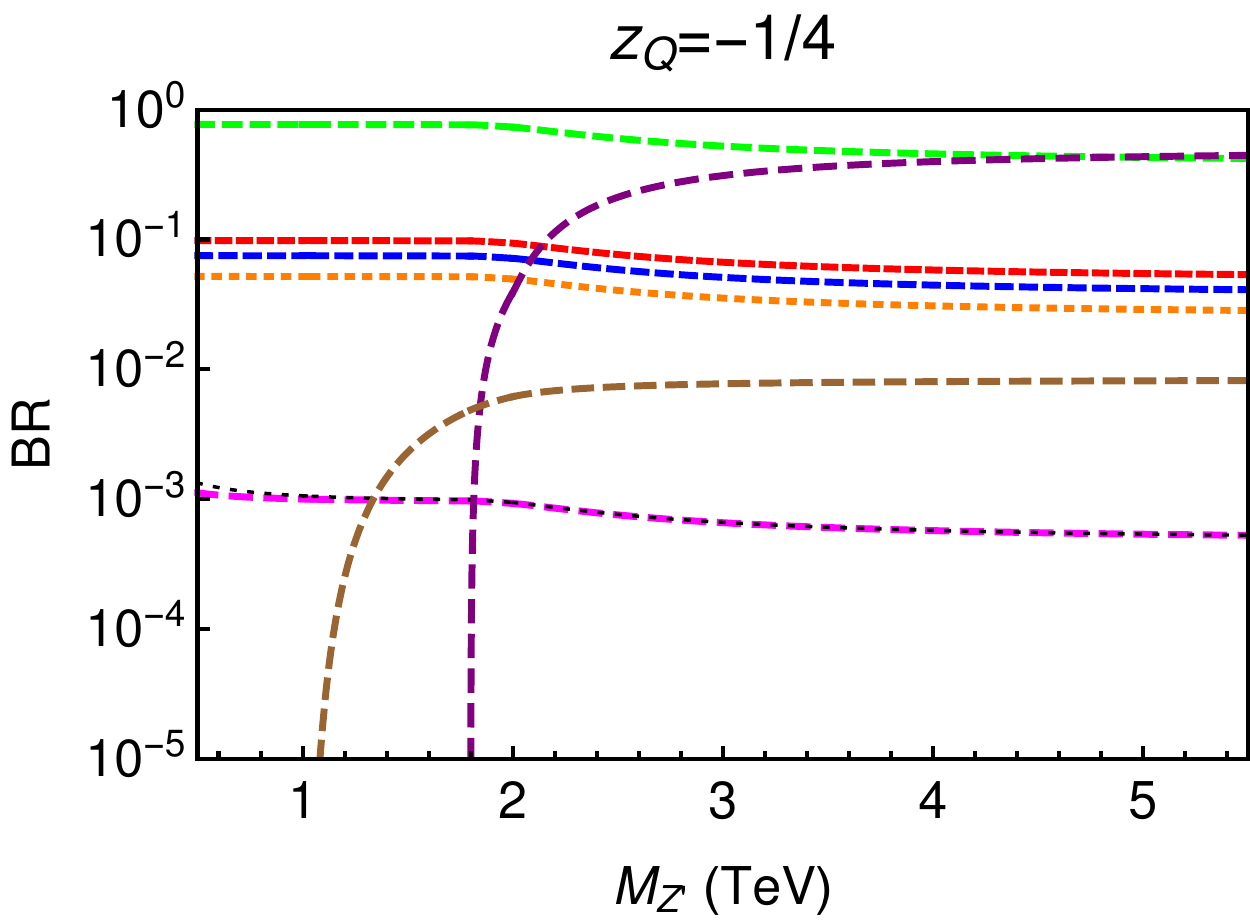}\label{fig:ZpBRa}}
\subfloat[]{\includegraphics[height=5.5cm,width=5.2cm]{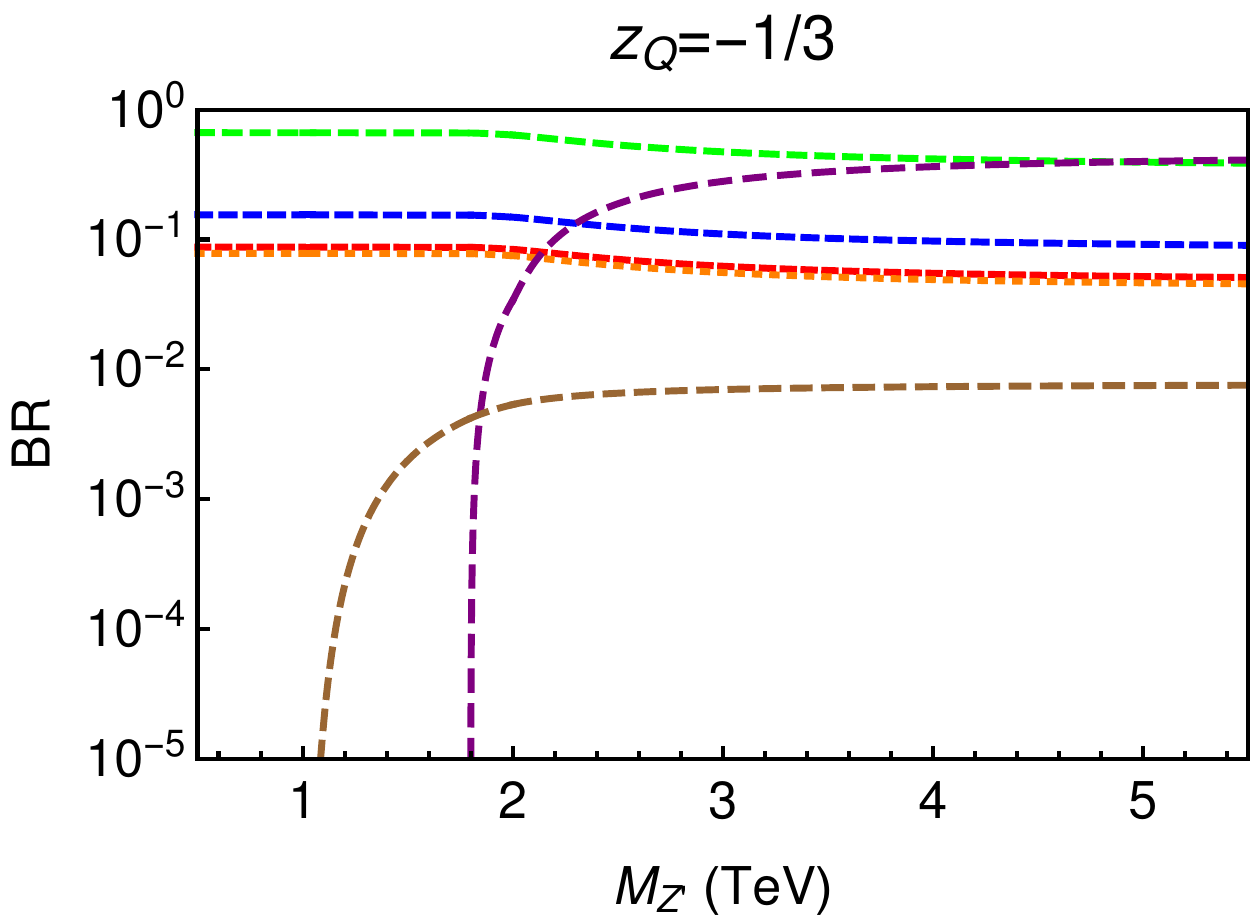}\label{fig:ZpBRb}}
\subfloat[]{\includegraphics[height=5.2cm,width=5.2cm]{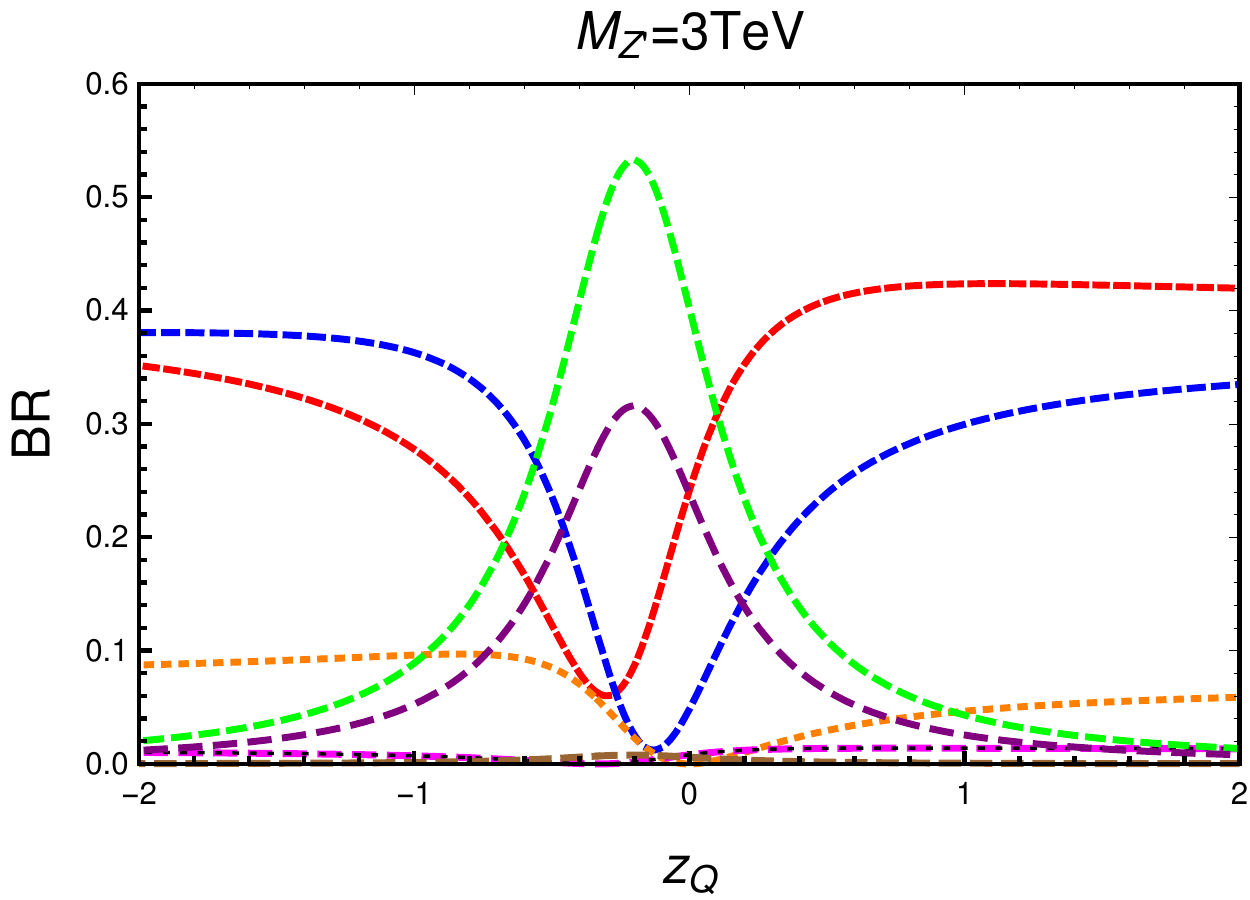}\label{fig:ZpBRc}}\\
% \\
\includegraphics[scale=0.8]{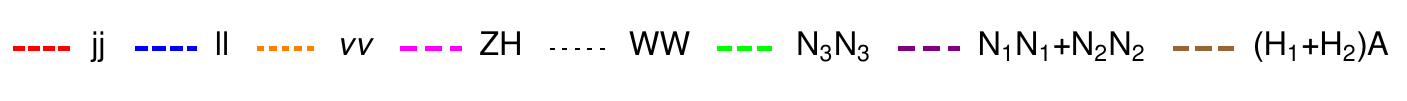}
% \subfloat[]{\includegraphics[height=6.5cm,width=7.5cm]{Figs/BR-NN_LL_vv_HA}\label{fig:ZpBRa}}\\
\caption{Branching ratios of various two-body decay modes of $Z'$ as functions of its mass $M_{Z'}$ for (a) $z_Q=-1/4$ and (b) $z_Q=-1/3$. In (c), we show similar BRs as functions of $z_Q$ for $M_{Z'}=3$ TeV. For these plots, we choose $g_z=0.15$. Here, $j$ includes $u,d,c,s,b$ and $\ell$ includes $e,\mu,\tau$.}
\label{fig:ZpBR}
\end{figure}

As a particular benchmark point in the parameter space, we consider 
\begin{equation}
M_{Z'} = 3 \rm TeV \ , \qquad M_{N_1}, M_{N_2}\sim 1\rm TeV \ ,
\qquad M_{H_1}, M_{H_2}\sim 1\rm TeV \ .
\end{equation}
These values are chosen so as to ensure that all possible two-body
decay modes are kinematically open. In Figs.~\ref{fig:ZpBRa} and \ref{fig:ZpBRb}, we
display the BRs of various decay modes of $Z'$ as a function of
$M_{Z'}$ and for two particular values of $z_Q$.  The kinks in these
figures are easily understood as manifestations of new thresholds
opening up. Inclusion of off-shell (three-body or even four-body final
states) serves to smoothen out the kinks.

Given the large $\mathrm{U}(1)_z$ charge of the $N_3$ and its small
mass, it is understandable that the $Z'\to \bar N_3 N_3$ mode
overwhelmingly dominates. While, individually, the $Z' \to \bar
N_{1,2} N_{1,2}$ are expected to be suppressed (compared to the $\bar
N_3 N_3$ one) by only a factor of $16/25$, these also suffer an
additional kinematic suppression, especially at lower values of
$M_{Z'}$. For high $Z'$ masses, together, these would slightly
overcome the former mode. Of the decays into exclusively SM decay
modes, of particular interest (since the corresponding SM backgrounds
are not too large) are the ones to $W^+ W^-$ and $Z H_{\rm SM}$. Equal
on account of the Goldstone equivalence theorem, these tend to be
small as their amplitudes are proportional to the $Z\leftrightarrow
Z'$ mixing. In particular, for $z_Q= -1/3$, the coupling of the SM
Higgs $H$ to the $Z'$ vanishes identically (at least at the tree
level), and so does the $Z$-$Z'$ mixing and, hence, these modes
vanish too\footnote{This also has immediate
  impact on the low-energy observables, especially those measured at
  the $Z$-peak, and is reminiscent of the $B-L$ model as discussed,
  for example, in Ref.~\citep{Ekstedt:2016wyi}.}. Less suppressed are
the modes $Z'\to (H_1+H_2)A$.

That the decays into exclusively SM modes tend to be subdominant for
$|z_Q| \lapp 1/3$ is but a reflection of the charges, which also
accounts for the ratios of the said partial widths. This suppression,
though, does not hold for larger values of $|z_Q|$, as is reflected by
Fig.~\ref{fig:ZpBRc}. Indeed, for a sufficiently large $|z_Q|$, it is
the dilepton or dijet modes that dominate. It is interesting to note
that the value of $z_Q$ that minimizes $Z'$ production is not the same
as when these decay modes minimize. While we could have chosen a $z_Q$
such that, say, the dilepton signal at the LHC is minimized
(rendering the model relatively free from constraints), we choose not
to do so.

It is also worthwhile to consider the decay modes of $N_{1,2}$. The
exact details would, of course, depend on the specific pattern of the
Yukawa couplings. The decay to $W^\pm$ and $\ell^\pm$ is higher than $50\%$ 
in this case because of the presence of extra $U_{PMNS}$ factor in the decays to $Z/H$ and $\nu$. In calculating the decay widths, as shown in appendix, we have assumed that $M_{Z'} \gg M_N \gg M_{h^0}$.

\subsection{Exclusion limits}

Once the $Z'$ has been produced at a collider, it can be detected only
through its decays. The leading contribution, at the LHC, accrues from
$q\bar q$ fusion and, for moderately large $|z_Q|$, grows as
$z_Q^2$. Thus, a large $z_Q$ would facilitate detection and we
deliberately choose to eschew this part of the parameter space, considering
instead the case of
moderately small $z_Q$ values, when the production cross sections are not
too large. Consequently, one needs to concentrate on decay modes that are not
highly suppressed.  As we have seen, for much of the parameter space
of interest, $Z' \to \bar N_3 N_3$ is the dominant decay mode.  The
$N_3$ is not only very light, but also has a highly suppressed
coupling to lighter species (the SM-like neutrinos) and, consequently,
does not decay within the detector. Thus, this mode is not directly
visible. However, with the emission of a visible particle ({\em e.g.},
$q g \to q Z' \to q \bar N_3 N_3$), one could, instead have a signal
comprising of a single jet accompanied by missing transverse
momentum\footnote{Similarly, monophoton, mono-$Z$ or mono-$W$
    signals (accompanied, in each instance by a transverse momentum
    imbalance) are possible too, but these suffer from additional
    coupling constant suppressions.}. Although the SM backgrounds 
to this final state is well studied, the sheer size of the background and the
paucity of kinematic variables to play with renders this mode a relatively 
insensitive probe for a very heavy $Z'$.

As Fig.~\ref{fig:ZpBR} shows, for such $z_Q$, the $\bar N_\alpha N_\al
\, (\alpha = 1,2)$ modes, together, can be competitive with the $\bar
N_3 N_3$ mode. With the $N_\alpha$ decay branching fractions
essentially being given by Eq.~\eqref{Ndecay}, a variety of final
states are possible. Particularly intriguing is the possibility of the
same-sign dilepton final state, that may arise when both the $N$'s
decay into the same sign charged lepton, with the $W$'s subsequently
decaying into, say, jets. Also possible are the trilepton plus jets
and the four-lepton final states, albeit with smaller cross
sections. Many of these have been studied extensively, not only in the
context of many popular new physics scenarios (such as supersymmetry
or theories defined in higher dimensions), but also in those similar
to
ours~\cite{Das:2017pvt,Das:2017flq,Das:2017deo,Deppisch:2019kvs,Das:2019fee,Chiang:2019ajm,Das:2018usr,Das:2017nvm}. A
simple scaling of the cross sections convinces one that the parameter
space required for reproducing the neutrino masses and mixings would be accessible only once the high-luminosity version
of the LHC is operational. Rather than delve into the
details thereof, we concentrate instead on the most sensitive probe.

\subsubsection{From dilepton and dijet data}

Despite the relatively smaller branching fraction, the decay of the
$Z'$ to a pair of charged leptons provides the strongest constraints
on the parameter space, followed by the dijet signal.

Since $\Gamma(Z') \ll M_{Z'}$, expressing dilepton (or dijet)
production in terms of an on-shell $Z'$ production followed by its
decay constitutes an excellent approximation. The leading order
contribution emanates from $q\bar q$ fusion, and has a simple
structure as given in Eq.~\eqref{cs_Zpr_prodn}. We, though, include the
next-to-leading order QCD corrections, parametrizable in terms of a
$K$-factor of 1.3~\cite{Gumus:2006mxa}. As for the parton fluxes, we
use the NNPDF2.3LO~\cite{Ball:2012cx} parton distributions, with the
natural choice for the renormalization and factorizations scales,
namely $\mu_F = \mu_R = M_{Z'}$.

Exclusion bounds on the model parameters, from a given experiment, can
be obtained by comparing the expected signal strength with the upper
bound (UB) on new physics events that the non-observation of an excess
in the said experiment implies. To this end, we use the
dilepton~\cite{Aad:2019fac,CMS:2019tbu} and
dijet~\cite{Aad:2019hjw,Sirunyan:2019vgj} resonance search data that
the two LHC experiments have collected (at $\sqrt{s} = 13$ TeV) with
an approximate integrated luminosity of 140 fb$^{-1}$. We start by
summarizing the experimental results:

\begin{itemize}
\item \textbf{ATLAS dilepton}~\cite{Aad:2019fac}: The ATLAS
  collaboration has performed a high-mass spin-1 resonance search in
  the dilepton final state in the mass range of $0.25$ TeV to 6 TeV
  with an integrated luminosity of 139 fb$^{-1}$. We recast their
  upper bound on fiducial $\sg\times BR$ for a spin-1 selection with a
  width/mass hypothesis of 1.2\% as is applicable for our
  analysis\footnote{Ref.~\cite{Aad:2019fac} has performed the
      analysis for several values of this ratio, and we choose the one
      closest to our situation. We have checked that our conclusions
      are not too sensitive to this choice.}. The definition of the
  fiducial phase-space region and the fiducial selection efficiency
  can be found in Ref.~\cite{Aad:2019fac}. While this efficiency differs
  slightly for the dielectron and dimuon channels and varies, in
  addition, with the resonance mass, for the sake of simplicity, we
  use a fixed fiducial selection efficiency of $0.6$ for the entire
  dilepton invariant mass range\footnote{We have checked that the
    consequences of this approximation are too small to be relevant.}.
  We obtain the observed $\sg\times BR$ UB from the
  \textsc{HepData}~\cite{Maguire:2017ypu} repository.

\item \textbf{ATLAS dijet}~\cite{Aad:2019hjw}: 
For the dijet channel (also done with the same luminosity),
 the collaboration presents an UB on $\sg\times
BR\times A$ (where the acceptance $A$ 
can be approximated to 0.4). Recasting the data presented in
 Fig. 8a of~\cite{Aad:2019hjw} for a generic Gaussian signal
in the inclusive channel with a 3\% width/mass hypothesis, we obtain our 
exclusion limits.

\item \textbf{CMS dilepton}~\cite{CMS:2019tbu}: The CMS collaboration
  has performed a high-mass spin-1 resonance search in the dilepton
  final state in the mass range from $0.2$ TeV to $5.4$ TeV.  Working
  with an integrated luminosity $\sim 140$ fb$^{-1}$, they present an
  upper bound on $\sg\times BR$, assuming the SM value for the
  width/mass ratio, namely $0.6\%$.

\item \textbf{CMS dijet}~\cite{Sirunyan:2019vgj}: For the high-mass
  dijet events, the collaboration uses an integrated luminosity of 137
  fb$^{-1}$. We recast the observed UB on $\sg\times BR\times A$ (with
  $A=0.5$) taken from Fig.~10 (lower panel) of
    Ref.\cite{Sirunyan:2019vgj} for the spin-1 resonance with a
  width/mass ratio of 1\%. In all these four searches discussed above,
  we use the ones with smallest width/mass ratio which are available
  in those analyses as in our case $Z'$ width is much smaller compared
  to its mass. 
\end{itemize}
\begin{figure}
\centering
\includegraphics[scale=1]{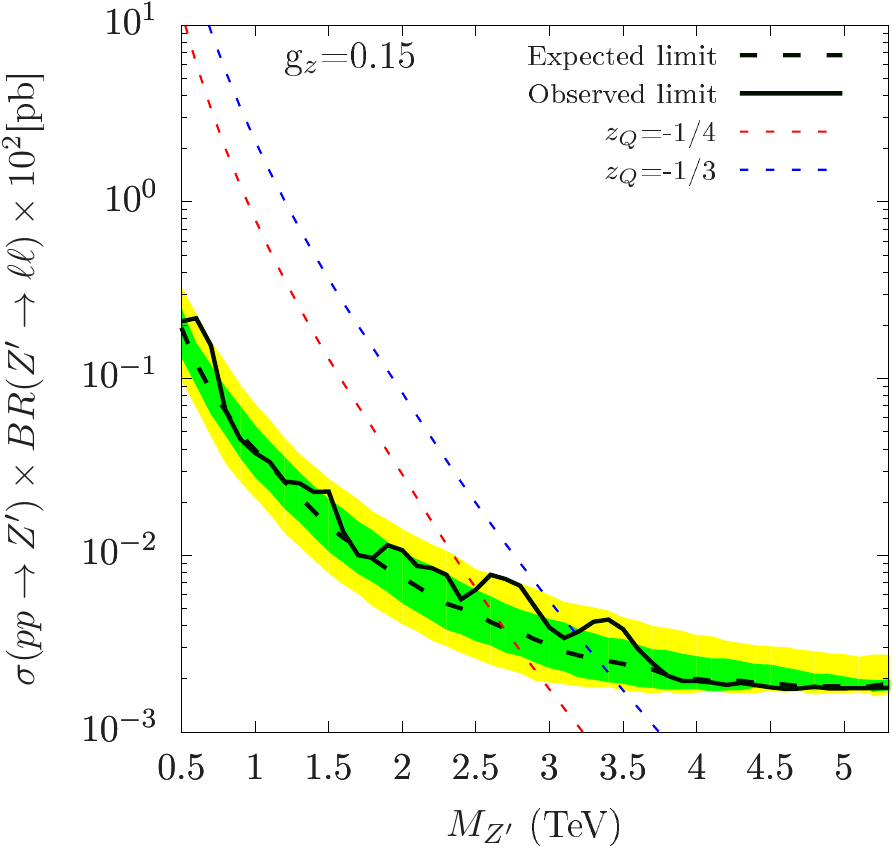} 
\caption{Comparison of the $95\%$ CL upper bound on the observed and the
  expected $\sg(pp\to Z')\times BR(Z'\to\ell\ell)$ obtained from the
  ATLAS dilepton resonance search data at the 13 TeV LHC with $L=139$
  fb$^{-1}$ with the theoretical predictions of our model for
  $z_Q=-1/4$ and $-1/3$ choices. We use the reference value for the
  $\mathrm{U}(1)_z$ gauge coupling $g_z=0.15$. The green and yellow
  bands represent the $1\sg$ and $2\sg$ uncertainty regions of the
  expected values respectively.}
\label{fig:Zpmassex}
\end{figure}

As a particular example, we display, in Fig.~\ref{fig:Zpmassex}, the
exclusion limits on the $Z'$ mass, as gleaned from the ATLAS dilepton
resonance search data, for two choices of $z_Q$, namely $z_Q=-1/4$ and
$-1/3$. Working with the fiducial $\sg_{fid}\times BR$ as provided in
the ATLAS paper~\cite{Aad:2019fac}, we obtain $\sg\times BR$ using a
fiducial selection efficiency of $0.6$.  The lower limits on $M_{Z'}$,
for $g_z=0.1$, are about 2.2 (2.6) TeV for $z_Q=-1/4 \,
(-1/3)$. 
%\sout{Note that since the dilepton BR is minimized for
%  $z_Q=-1/4$}, \comment{not really!}  
  The difference in the two exclusions can be traced to two factors,
  namely a slightly smaller $Z'$-production cross section, as well as
  a slightly smaller branching fraction into a charged lepton
  pair. Note that (as promised earlier) the dependence on $z_Q$ is not
  too severe.  Were one to be interested in the mass exclusions for
  other $z_Q$ and $g_z$ values, these could be obtained trivially by
  realizing that the production cross section scales as $g_z^2$, and
  reading off the dilepton BR from Fig.~\ref{fig:ZpBR}.

\subsubsection{Low-energy observables}

A nonzero value of $z_H$, the
$\mathrm{U}(1)_z$ charge of the SM Higgs doublet, induces
tree-level $Z\leftrightarrow Z'$ mixing. This has two main ramifications:

\begin{itemize}
\item Tree-level contributions to the oblique parameters are
  induced. In particular, the tree-level contribution to the
  $T$-parameter is given by~\citep{Appelquist:2002mw},
\begin{align}
\alpha_{EM}T^{\rm new} = \frac{\Pi_{ZZ}^{\rm new}}{M_Z^2} = \frac{M_Z^2 - (M_Z^0)^2}{M_Z^2}.
\end{align}
Here, $M_Z$ is the $Z$-boson mass in the new theory and, for our
purposes, it suffices to consider the tree-level expression as given
in Eq.~\eqref{eq:mass}. Similarly, $M_Z^0$, the mass within the SM, is
given by $M_Z^0 = g_w v_h/2\cos w$ at the tree-level.  And, finally,
$\al_{EM}$ denotes the fine-structure constant at $Z$-pole. In
effecting the actual calculation, the higher-order SM contributions
would, of course, have to be taken into account, and we have done so.
On the other hand, loop corrections to $T$ wrought by new physics are
further suppressed by large masses and the $Z\leftrightarrow Z'$
mixing angle and can be safely neglected.  We use the value
$T=0.07\pm 0.12$ Ref. \cite{Tanabashi:2018oca} in our analysis. 

\item A related constraint arises from the measurement of the
  $Z$-coupling of the light fermions, occasioned, again, primarily by
  the $Z\leftrightarrow Z'$ mixing.  Determined from the
  forward-backward asymmetry or through the line-shape of the
  $Z$-resonance, these observables as also the
  $Z$-width~\cite{Appelquist:2002mw} are very precisely
  measured~\cite{Tanabashi:2018oca}.
  
\item Another relevant constraint can come from the LEP
  measurements. The $Z'$ boson, despite being heavier than the
  LEP energies, can contribute to the $e^+e^-\to \bar{f}f$ processes
  through the interference with the $\gamma$ and $Z$ mediated
  processes. For the sequential-SM, the 95\% confidence level lower
  limit on the $Z'$ mass is 1760 GeV as obtained from the LEP
  data~\cite{Schael:2013ita}. In our case, this limit 
  is much more relaxed to the point of being irrelevant 
  since the $Z'$ couplings with leptons
  and quarks are much smaller. than the SM-like couplings for our
  benchmark parameters. Therefore, we do not consider LEP constraints
  in our analysis.
\end{itemize}

Once again, due to the vanishing of $z_H$ for $z_Q=-1/3$, there is no
tree-level $Z\leftrightarrow Z'$ mixing, and neither of the
aforementioned constraints are applicable, at least at the
tree-level. The one-loop effect is too small to be of any consequence.
As for $z_Q=-1/4$, our choice commensurate with neutrino phenomenology
as well as possible unification, owing to it not being too far from
$-1/3$, the $Z\leftrightarrow Z'$ mixing is still very small, and 
such low-energy observables do not strongly constrain the
parameter space.

\begin{figure}[!h]
\centering
\subfloat[]{\includegraphics[scale=.5]{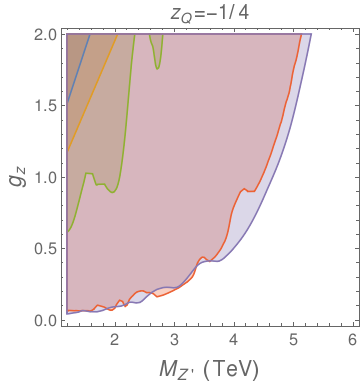}\label{fig:Zpregexa}}
\subfloat[]{\includegraphics[scale=.5]{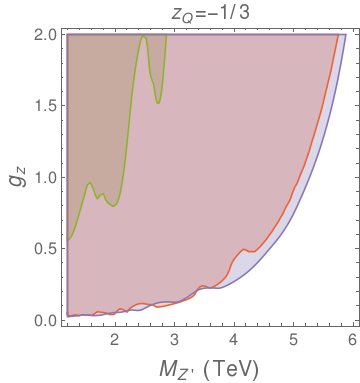}\label{fig:Zpregexb}}\\
\subfloat[]{\includegraphics[scale=.5]{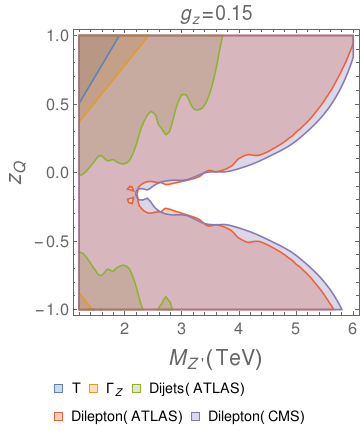}\label{fig:Zpregexc}}
\caption{Exclusion regions in the $M_{Z'}-g_z$ plane for fixed (a) $z_Q = -1/4$ and (b) $z_Q = -1/3$ and in (c) in the $M_{Z'}-z_Q$ plane for fixed $g_z=0.15$. We show exclusion regions using $T$-parameter, $Z$-width, and the latest dilepton and dijet data from the LHC.}
\label{fig:Zpregex}
\end{figure}

We show the exclusion plots in the $M_{Z'}-g_z$ plane in
Figs.~\ref{fig:Zpregexa} and \ref{fig:Zpregexb} for $z_Q=-1/4$ and
$z_Q=-1/3$ choices respectively, while in 
Fig.~\ref{fig:Zpregexc}, we show the similar exclusion in the
$M_{Z'}-z_Q$ plane for fixed $g_z=0.15$. In each case, we show exclusion regions using the latest dilepton, dijet
resonance search data from the LHC, as well as those coming from the
$T$-parameter and $Z$-width measurements.  As expected, the dilepton data does impose severe
constraints. However, owing to $Br(Z' \to \bar\ell \ell)$ assuming its minimum at
  $z_Q \sim -1/4$, the lower limit on $M_{Z'}$ for both values of $z_Q$ considered here is as
low as $\approx 2$ TeV for $g_z\sim 0.15$. And while the
dijet branching fraction is comparable
to the dilepton one, this mode suffers from a much larger (QCD) 
background, and consequently the bounds are much weaker. As
for $T$-parameter and $Z$-width measurements, since
tree-level $Z\leftrightarrow Z'$ mixing is absent for $z_Q=-1/3$,
the constraints are virtually nonexistent, and continue to be
 very weak for $z_Q=-1/4$ as well.

\section{A Dark Matter candidate?}
\label{sec:DM}

As we have already seen in Sec.~\ref{sec:u1_extn} and
Sec.~\ref{sec:numass}, the model may have three ultralight particles,
{\em viz.}
\begin{itemize}
\item a pseudoscalar pseudo-Nambu-Goldstone boson $A$, 
\item a largely singlet neutrino $N_{3R}$, and 
\item an even lighter doublet-like neutrino $\nu$. 
\end{itemize}
The lightest of the three, being stable, would be an apparent candidate 
for the dark matter. Indeed, given the lifetimes and interaction strengths, 
the ``dark sector'' could be much richer. 

It is, of course, well-known that a light doublet-like neutrino would
have decoupled while still relativistic and would constitute hot dark
matter. Since this would have interfered with large-scale structure
formation, a hot dark matter can constitute only a minor fraction of 
the total dark matter relic density and, fortunately, the very structure 
of the SM ensures that the usual neutrinos satisfy this condition. And
with $\nu$ being overwhelmingly doublet-like, it too would automatically 
satisfy the same. 

This allows us to direct our attention to the pseudoscalar $A$ and the
singlet-like neutrino $N_{3R}$. As we have already discussed, the
former's mass can be trivially uplifted by the dint of adding a third
singlet $\chi_3$ such that nontrivial trilinear and/or quadrilinear
terms are admissible in the scalar potential. While either of soft and
hard breaking of the global symmetry---down to a single $\mathrm{U}(1)$ which
is gauged---will render the pseudoscalar(s) massive, in the case of
the former (soft breaking), the resultant mass is controllable (and we
have a pNGB). This constitutes a particularly simple strategy as the
neutrino masses and mixings are essentially left unchanged. As for
collider signals, while the partial widths $Z' \to A \xi_i$ would be
altered on account of the $A$ picking up a mass, the changes are not
very significant for relatively small $M_A$ values. More importantly
though, the $A$ may now decay within the detector, thereby eliminating
this particular source of a missing transverse energy signal. This,
however, would be replaced by more exotic (and, hence, more visible)
signals at the LHC, perhaps including displaced vertices. 

 Concentrating on the $N_{3R}$(--like) state, let us begin by examining
its mass. As a perusal of Eq.~\eqref{nu_mass_our} shows, the largest
contribution to this mass would accrue from the seesaw like mechanism
involving the $N_{1R}$ and $N_{2R}$ fields. This would lead to an
effective mass for the mass-eigenstate $\Psi$ of\footnote{Note that
  $m(\Psi)$ is still much larger than the $s_{33}$ term in
  Eq.~\eqref{nu_mass_our}, a situation very analogous to that in the
  doublet-sector, namely the large difference between $M_{3 \times
    3}$ and the corresponding Weinberg operator.}
\[
  m(\Psi) \sim \left (s_{\alpha 3} \frac{x_1^4 x_2}{\Lambda^4}\right)^2 \; 
         \frac{\Lambda}{x_2^2}
         \sim s_{\alpha 3}^2 \xi^7  x \ .
\]
For $s_{\alpha 3}$ of around $0.05$ and the two heavy RHNs of the
order $1.2 ~ \rm TeV$, we get the mass of $N_3$ to be a few keVs. A
sterile neutrino at such a scale is particularly interesting from the
dark matter perspective because it can have a very long lifetime,
comparable to the age of the Universe.  For the keV scale DM
candidates, almost the entire observed relic density can 
be accounted for only if $m_{DM} > 0.4 ~ \rm
keV$~\cite{PhysRevLett.42.407,Adhikari:2016bei}, which seems plausible
in our case. Still, large mixings with
  neutrinos, while being a desirable feature from the
collider search point of view, can end up producing way more
dark matter than we require, thereby
potentially spoiling its DM candidature.

A state as light as this can only have two types of decays, namely
$\Psi \to 3 \nu$ (where $\nu$ are the SM-like mass eigenstates) and
$\Psi \to \nu_i + \gamma$. Clearly, the latter are loop-suppressed
and, hence, of relatively little significance. As for the former
set of modes, it is instructive to consider, formally, the
lowest-dimensional operator within the effective theory that can lead
to them. This is most easily done in terms of the gauge eigenstates,
{\em viz.},
\[
 \frac{a_1}{\Lambda^4} \, 
         (\bar \ell_i \gamma_\mu \ell_j) \, 
(\bar \ell_k \gamma^\mu N^c_{3R}) H^* \chi_2 \ .
\]   
For $a_1 \sim {\cal O}(1)$, this leads to 
\[
         \tau_\Psi \gapp \frac{10^{25}}{a_1^2} \, {\rm s}  \times 
            \left(\frac{100 \, {\rm MeV}}{m_\Psi}\right)^5 ,
      \]
which is comfortably larger than the age of the universe ( $ \tau_U \sim 5
\times 10^{17}$ s). 
 Naively, it might seem then that we might have 
as well admitted a different structure that would have allowed 
for a much heavier $N_{3R}$.  

This, however, is misleading. As we have seen in Sec.\ref{sec:numass},
the diagonalization of the mass matrix
leads to mixings between the $\nu$-like and the $N_3$ like eigenstates. 
In other words, the mass eigenstates $\Psi, \nu'$ are symbolically 
given by
\[
\Psi \approx \cos\theta_i N_3 + \sin\theta_i \nu_i \ , \qquad 
\nu' \approx -\sin\theta_i N_3 + \cos\theta_i \nu \ ,
\]
where generational dependence has been omitted. Other mixings
connecting $\nu_i, N_{3R}$ with $N_{1,2}$ are not taken into account
as they are much suppressed and, therefore, not significant
for the issue at hand.

Owing to the fact that the $N_{3R}$ is a singlet, this mixing
immediately leads to a $Z\bar \Psi\nu'$ coupling which, in turn, leads
to a $Z$-mediated contribution to the $\Psi \to 3 \nu$ decay. The
corresponding partial widths are
\begin{equation}
\Gamma_i = \Gamma(\Psi \to \nu_i \bar \nu_j \nu_j) \sim 
         \dfrac{G_F^2 M_{N_3}^5}{192 \pi^3} \sin^2 \theta_i \, 
            \left(1 - \frac{\delta_{ij}}{2}\right) \ ,
\end{equation}
leading to a lifetime $\tau_\Psi \gapp 10^{18}$s, for $m_N \sim 6$~keV
and mixing of $\mathcal{O}(10^{-2})$. So for such masses and mixings,
we still can barely have $\tau_\Psi > \tau_U$.  This ostensible
enhancement compared to the earlier estimate can be traced back to the
fact that several powers of $(x/\Lambda)$ in the aforementioned
effective operator are actually subsumed in the suppression of
$m_{N_3}$ itself, and must not be double counted.

At this point, several issues need to be delved upon. While it might
seem that we naturally have $\tau_\Psi > \tau_U$, in reality,
$\tau_\Psi$ depends crucially on the values of the Wilson
coefficients, and, thus, such a requirement on the lifetime imposes
conditions on the WCs.  It should be appreciated though that, for such
a light DM particle, $\tau_\Psi > \tau_U$ is not a strict
requirement. In fact, a $\tau_\Psi$ value even somewhat smaller than
$\tau_U$ may also be admissible for this would only mean that a
fraction of the DM has decayed in the course of the Universe's
evolution yet leaving behind sufficient relic density. Since the decay
is into neutrinos alone, the only discernible effect would be through
altering the neutrino-photon ratio in the early universe, thereby
altering the effective number of relativistic degrees of freedom, and
thus invite constraints from this measurement. However, the existence
of such a restriction depends crucially not only on the epoch of this
decay, but also on specific mechanism ({\em e.g.}, freeze-in versus
freeze-out) of DM relic density generation, and is not immediately
applicable to the case at hand.

On the other hand, even if a $\tau_\Psi$ value somewhat smaller than
$\tau_U$ is obtained, these same operators would lead, at one loop, to
$\Psi \to \nu_i + \gamma$ manifesting itself through X-ray lines. The
non-observation of such signal in different low energy experiments,
mandates~\cite{Boyarsky:2005us,Abazajian:2006jc} that the sterile-SM
neutrino mixing angle satisfies $\theta_i \ll 10^{-2}$ if a 6 keV
$\Psi$ were to provide full DM relic density. The simplest way to
satisfy such seemingly incompatible constraints would then be to
assume that the $\Psi$ provides for only a small fraction of the relic
density in the form of a potentially warm component, especially since
for certain regions of the parameter space, it could be produced
non-thermally. More interestingly, the $\Psi$ has considerable
self-interaction (mediated by the $Z'$) with a suppressed, but
long-distance component mediated by the pseudoscalar $A$. This has the
potential to provide some pressure to the DM fluid and, thereby,
address certain long-standing issues pertaining to details of
structure formation.

To anoint $\Psi$ to be the main or even a significant DM constituent,
one must ascertain not only whether the correct relic density can be
reached but also whether the scenario falls foul of other constraints,
both cosmological as well as those emanating from laboratory tests
(both direct and indirect detection). This demands detailed analysis
that is beyond the scope of the present work.  However, at the same
time, we want to emphasize a few general issues. In our quantitative
analysis we made several simplifying assumptions regarding $x_i,
\Lambda$ and some of the WCs. Tweaking these assumptions can
substantially change the masses and mixings of the neutrino, as is
seen in the context of a Frogatt-Nielsen scenario through the
introduction of different scales through different powers of a scaling
factor~\cite{Kamikado:2008jx}. In a similar vein, by altering the
$\mathrm{U}(1)_z$ charges of the $N_i$ (while maintaining anomaly cancellation)
and/or the scalar fields $\chi_a$, the neutrino mass matrix can be
changed. This would allow us a much larger mass for the $N_3$-like
state, {\em viz.} ${\cal O}({\rm MeV})$, with further suppressed
$N_3$--$\nu_i$ mixings thereby still allowing for $\tau_\Psi \gapp
\tau_U$. Consequently, the standard freeze-in mechanism would hold for
such a DM.  We would like to postpone these issues to a future
project.

To examine the falsifiability of our hypothesis, it is important to
consider the strength of the interactions that mediate low-energy
scattering involving $\Psi$ and the SM particles. Fortunately, such
interactions are not unduly suppressed thanks to the fact that both
$N_{3R}$ and the SM fermion carry $\mathrm{U}(1)_z$ quantum
numbers. Consequently, the interaction strength is governed only by
$g_z^2 / M_{Z'}^2$ or, equivalently, by $x_i^{-2}$, and in the case of
freeze in DM generation, relic density is proportional to this
interaction, that, $M_{Z^{\prime}}$ being at the TeV-scale, exactly
represents an example of how entire relic density can be reproduced.
Rather than present a full analysis, we refer the reader to the
existing literature.  For example, it has been shown, in an analogous
context, in Ref.\cite{Choudhury:2019tss} that a parameter space,
consistent simultaneously with the requisite relic density, the
measurement of the cosmologically relevant effective relativistic
degrees of freedom and energy injection, from DM annihilation, into
the cosmic microwave background radiation, can be found. The required
interaction strength is of the same order as what transpires naturally
in our model and would leave such particles undetectable in the
currently operative (satellite-based) indirect detection
experiments~\cite{Choudhury:2019tss}.  Even more interestingly, such a
DM is likely to be detectable not only at the next generation of
direct detection experiments, but also at the Super-Belle
detector~\cite{Choudhury:2019sxt}.

Before we end this section, we would like to remind the reader of a
possibility that we did not elaborate on.  Consider, for example, the
case where the extra global $\mathrm{U}(1)$ is not broken by terms in the
potential. The Yukawa couplings, nonetheless, do break it and,
consequently, quantum corrections would lift the mass of the Goldstone
by a tiny amount, leaving it stable on cosmological time scales. Free
from restrictions (such as those imposed by X-ray or Lyman-$\alpha$
observations), this could, again, play a significant role in the
evolution of galaxy clusters etc.~\cite{Arina:2019tib}. A detailed examination of
such effects is beyond the scope of this paper and is postponed for a
future study.

\section{Summary and Conclusion}
    \label{sec:sumcon}

With the aim of explaining neutrino masses without invoking either
ultrasmall Yukawa couplings or an almost inaccessible new (seesaw)
scale, we consider a scenario where the gauge symmetry has been
augmented by an extra $\mathrm{U}(1)_z$. If its action on the SM particles is
nontrivial, but generation-invariant (so as to allow for a single SM
Higgs to give masses to the charged leptons), then the possible charge
assignment for the right-handed neutrinos (RHN) is severely restricted
by the requirement of gauge (and mixed gauge-gravity) anomaly
cancellation. (We assume here that, unlike in certain popular schemes
such as the inverse seesaw mechanism, we have the minimum possible
number of RHNs.)  Only the most trivial such assignment allows for
tree-level neutrino Dirac mass terms. On the other hand, bare Majorana
mass terms cannot be incorporated. Indeed, analogues of the Weinberg
term can be written only if the new Higgs breaking the $\mathrm{U}(1)_z$ have
one of two specific choices of the charge.

For any choice of the RHN charges other than the most trivial one, not
only are renormalizable Dirac mass terms disallowed, but so are the
Majorana mass terms except for specific choices of the $\mathrm{U}(1)_z$
breaking Higgs bosons. Completely unrelated to this, the absence of
any resonance in the LHC data has pushed the mass of the new gauge
boson $Z'$ to above several TeVs. 

In view of this, we assume an agnostic standpoint claiming that any
such theory can, at best, be the low-energy limit of a more
fundamental theory, characterised by a cut-off scale $\Lambda$. This,
immediately, allows us to write non-renormalizable terms suppressed by
powers of $\Lambda$. While a wide variety of such terms, in principle,
can be written, we concern ourselves only with the neutrino
sector. Invoking the next to the trivial quantum number assignment for
the RHNs, we then write down all relevant higher-dimensional terms
{\em \'a la} the Froggatt-Nielsen mechanism. Using the power of higher
dimensional operators to the hilt, we generate tiny neutrino masses
without any need to invoke tiny Yukawa couplings. Indeed, even without
using all the free parameters of the theory, it can naturally
reproduce the experimentally observed neutrino mixings and
mass-squared differences, while satisfying the cosmological bound on
the sum of masses as well as that from non-observation of neutrinoless
double-beta-decay. Simultaneously, it prophesies, amongst others
  \begin{itemize}
    \item a pair of heavy RHNs $N_{1,2}$ at the 1 TeV mass scale that decay 
      promptly into $\ell W$, $\nu Z$ and promise interesting 
      signals at the high-luminosity run of the LHC;
    \item a moderately heavy $Z'$ ($m_{Z'} \lapp 3$ TeV for $g_{Z'}
      \sim 0.15$) that escapes LHC bounds---from dilepton and dijet
      searches---despite having unsuppressed couplings with the quarks
      and leptons, simply by virtue of decaying primarily into the RHNs.
      Similarly, for natural choices of $\mathrm{U}(1)_z$ charges (especially 
      those commensurate with possible charge quantization), the LEP 
      constraints such as those on the oblique parameters are trivially 
      satisfied;
    \item a light RHN $N_3$ in the keV---MeV range. With the $Z'$
      having a large branching fraction into a $N_3$-pair, and with
      the $N_3$ being stable at the collider timescales, this would
      lead to additional contribution to the monojet (monophoton) plus
      missing transverse momentum signal at the LHC; Indeed, for a
      large part of the parameter space, the $N_3$ can have a lifetime
      comparable to or even greater than that of the Universe and,
      thus, can constitute a warm DM component.
    \item a pseudoscalar pseudo-Nambu Goldstone boson, with its mass
      uplifted only by quantum corrections or additional soft terms in
      the scalar potential (the latter being absent in the simplest
      realization). This has the potential of being an additional
      contributor to the DM relic density (while escaping many of the
      constraints applicable to $N_3$). Furthermore not only does it
      have non-negligible self-interaction, but it can also mediate $N_3$
      scattering thereby playing an important in not only determining 
      the relic density, but also in engendering a non-negligible 
      pressure term for the DM fluid and thereby affecting the details 
      of structure formation.
\end{itemize}

The model presented, thus, offers much more than an understanding 
neutrino phenomenology. Not only does it offer tantalizing prospects 
at the LHC, but also intriguing avenues to explore in the context of 
dark matter and details of structure formation. We hope to return 
to more in-depth study of these issues in a future publication.

\section*{Acknowledgments}

DC and TM acknowledge partial support from the SERB, India under research
grant CRG/2018/004889. DC also acknowledges the European Union’s Horizon 2020 research and innovation program under Marie Sk{\l}odowska-Curie grant No
690575. KD acknowledges Council for Scientific and Industrial Research(CSIR), India for JRF fellowship with award letter no. 09/045(1654)/2019-EMR-1. SS thanks UGC for the DS Kothari postdoctoral fellowship grant with award letter No.F.4-2/2006 (BSR)/PH/17-18/0126.

\newpage
\section{Appendix}
\label{app:PW}

\subsection{Decay widths of $Z^\prime$}

In this Appendix, we provide the analytical expressions of the tree-level partial widths of various two-body decay modes of $Z'$. These expressions are computed using the Feynman rules obtained from the interaction Lagrangian shown before them.
\begin{itemize}

\item \underline{$Z^\prime \to \bar{f}f$:} For the following interaction Lagrangian, 
\begin{equation}
\mathcal{L}_{Z^\prime \bar{f}f} = g_L \bar{f_L}\gamma^\mu f_L {Z\textprime}_\mu +  g_R \bar{f_R}\gamma^\mu f_R {Z^\prime}_\mu\ ,
\end{equation}
the expression for the $Z^\prime \to \bar{f}f$ partial width is given by
\begin{equation}
\Gamma_{(Z^\prime \to \bar{f}f)}= \dfrac{N_c M_{Z^\prime}}{24 \pi} \sqrt{1-\dfrac{4 M_f^2}{M^2_{Z^\prime}}}\lt[\lt(g_L^2+g_R^2\rt)\lt(1-\dfrac{M_f^2}{M_{Z^\prime}^2}\rt) + 6 g_Lg_R\dfrac{M_f^2}{M_{Z^\prime}^2}\rt].
\end{equation}
In the above, $g_L$ and $g_R$ are the left- and the right-handed couplings respectively, $M_f$ is the mass of the fermion $f$ and $N_c$ is the corresponding number of colors.

\item \underline{$Z^\prime \to \nu_R\nu_R$:} For the following interaction Lagrangian, 
\begin{equation}
\mathcal{L}_{Z^\prime \nu_R\nu_R}=g_\nu (\nu{_R^c})^T \gamma^\mu\nu_R Z^\prime_\mu\ ,
\end{equation}
the expression for the $Z^\prime \to \nu_R\nu_R$ partial width is given by
\begin{equation}
\Gamma_{(Z^\prime \to \nu_R\nu_R)}= \dfrac{M_{Z^\prime}}{24 \pi}{g_\nu}^2 \lt(1-\dfrac{4 M_{\nu_R}^2}{M_{Z^\prime}^2}\rt)^{3/2}.
\end{equation}
where $g_\nu$ is the coupling and $M_{\nu_R}$ is the mass of the RHN.

\item \underline{$Z^\prime \to W^+ W^-$:} For the following triple gauge boson interaction with strength $\lm_W$,
\begin{equation}
\mathcal{L}_{Z^\prime  W^+ W^-}\supset\lambda_W Z^\prime_\mu(p_1) W^+_\nu(p_2)  W^-_\rho(p_3)\ ,
\end{equation}
the expression for the $Z^\prime \to W^+ W^-$ partial width is given by
\begin{equation}
\Gamma_{(Z^\prime \to W^+ W^-)}= \dfrac{M^5_{Z^\prime}}{192 \pi M_W^4} \lambda^2_W \lt({1-\dfrac{4 M^2_W}{M^2_{Z^\prime}}}\rt)^{3/2} \lt(1+\dfrac{20 M^2_W}{M^2_{Z^\prime}}+\dfrac{12 M^4_W}{M^2_{Z^\prime}}\rt).
\end{equation}

\item \underline{$Z^\prime \to Z S$:} From the following interaction with dimensionful coupling strength $\mu_S$,
\begin{equation}
\mathcal{L}_{Z^\prime  Z S} = \mu_S Z^\prime_\mu Z^\mu S\ ,
\end{equation}
where $S$ is a $CP$ even scalar, the expression for the corresponding partial width is given by
\begin{equation}
\begin{aligned}
\Gamma_{(Z^\prime \to Z S)}= \dfrac{\mu^2_S M_{Z^\prime}}{192 \pi M_Z^4}  \lt({1-\dfrac{\lt(2M^2_S-10M^2_Z\rt)}{M^2_{Z^\prime}}}+\dfrac{\lt(M^2_S-M^2_Z\rt)^2}{M^4_{Z^\prime}}\rt) \\ \times \lt(1-\dfrac{2\lt(M^2_S+M^2_Z\rt)}{M^2_{Z^\prime}}+\dfrac{\lt(M^2_S-M^2_Z\rt)^2}{M^4_{Z^\prime}}\rt).
\end{aligned}
\end{equation}

\item \underline{$Z^\prime \to S A$:} From the following interaction with cubic coupling 
\begin{equation}
\mathcal{L}_{Z' S A}=g_p Z^\prime_\mu \partial^\mu S A\ ,
\end{equation}
where $S$ is a $CP$-even scalar and $A$ is a $CP$-odd scalar
\begin{equation}
\begin{aligned}
\Gamma_{(Z^\prime \to Z S)}= \dfrac{g^2_p}{12 \pi  M^5_{Z^\prime}}  \Big[\lt(M_S - M_A - M_{Z^\prime}\rt)\lt (M_S + M_A - M_{Z^\prime}\rt) \\ \lt(M_S - M_A + M_{Z^\prime}\rt) \lt(M_S + 
    M_A + M_{Z^\prime}\rt)\Big]^{3/2}
\end{aligned}
\end{equation}

\end{itemize}
\subsection{Decay widths of heavy RHN}
The heavy RHN decay modes are given by the Lagrangian of the form:
\beq
\mathcal{L}= -\dfrac{g_w}{\sqrt{2}}\bar{l}_L \gamma_\mu U_{\nu N} N W^\mu -\dfrac{g_w}{2 \cos w} \bar{\nu}\gamma_\mu U^\dagger_{\nu \nu} U_{\nu N} N Z^\mu - \dfrac{H}{v_h}M^{diag}_N U^{\dagger}_{\nu N} U_{\nu \nu} \ \nu + h.c
\eeq
The decay rates are then given by:
\beq
\Gamma(N_\alpha \to W^- l^+_i)=\Gamma(N_\alpha \to W^+ l^-_i)\approx \dfrac{g_w^2}{64 \pi M_W^2} M_N^3 |(U_{\nu N})_{i \alpha}|^2,
\eeq
\beq
\Gamma(N_\alpha \to Z \nu_i)\approx \Gamma(N_\alpha \to H \nu_i)\approx \dfrac{g_w^2}{64 \pi M_W^2} M_N^3 |(U^{\dagger}_{\nu \nu}U_{\nu N})_{i \alpha}|^2,
\label{Ndecay}
\eeq

\

where $U_{\nu \nu}$ is approximately the $PMNS$ matrix. $U_{\nu N}$ is the mixing between the light SM neutrinos and the heavy RHNs given by \cite{Kang:2015uoc} \cite{PhysRevD.80.073012}
\beq
U_{\nu N}  =  - {\cal D} (M_N^{diag})^{-1}.
\eeq

\newpage
\bibliographystyle{JHEPCust}
\bibliography{references.bib}

\end{document}